\documentclass[journal]{IEEEtran}
\usepackage[pdftex]{graphicx}
\usepackage{subfigure}
\usepackage{threeparttable}
\usepackage{dcolumn}
\usepackage{booktabs}
\usepackage{bm}
\usepackage{multirow}
\usepackage{xcolor}
\usepackage{cite}
\usepackage{amsmath}

\usepackage{algorithm}
\usepackage{algorithmicx}
\usepackage{algpseudocode}
\usepackage{tablefootnote}
\algblock{ParFor}{EndParFor}
\algnewcommand\algorithmicparfor{\textbf{parfor}}
\algnewcommand\algorithmicpardo{\textbf{do}}
\algnewcommand\algorithmicendparfor{\textbf{end\ parfor}}
\algrenewtext{ParFor}[1]{\algorithmicparfor\ #1\ \algorithmicpardo}
\algrenewtext{EndParFor}{\algorithmicendparfor}

\begin{document}
\title{Fast MPEG-CDVS Encoder with GPU-CPU Hybrid Computing}
\author{Ling-Yu~Duan,
        Wei~Sun,
        Xinfeng~Zhang,
        Shiqi~Wang,
        Jie~Chen,
        Jianxiong~Yin,
        Simon~See,
        Tiejun~Huang,
        Alex~C.~Kot,~\IEEEmembership{Fellow,~IEEE,}
        and~Wen~Gao,~\IEEEmembership{Fellow,~IEEE}% <-this % stops a space
\thanks{L.-Y. Duan, W. Sun, J. Chen, T. Huang, and W. Gao are with the School of Electronics Engineering and Computer Science, Institute of Digital Media, Peking University, Beijing 100871, China (e-mail: \{lingyu, weisun199508, cjie, tjhuang, wgao\}@pku.edu.cn).}% <-this % stops a space
\thanks{X. Zhang and Alex C. Kot are with the Rapid-Rich Object Search (ROSE) Lab, Nanyang Technological University, Singapore (e-mail: \{xfzhang, EACKOT\}@ntu.edu.sg).}% <-this % stops a space
\thanks{S. Wang is with the Department of Computer Science, City University of Hong Kong, Kowloon, Hong Kong (e-mail: shiqwang@cityu.edu.hk).}
\thanks{J. Yin, S. See are with the NVIDIA AI Tech. Centre (e-mail: \{jianxiongy, ssee\}@nvidia.com).}
\thanks{Ling-Yu Duan, Wei Sun and Xinfeng Zhang are joint first authors, and Ling-Yu Duan is the corresponding author.}
%\thanks{Manuscript received April 19, 2005; revised September 17, 2014.}
}

\markboth{Submitted to IEEE Transactions on Image Processing, May~2017}%
{Shell \MakeLowercase{\textit{et al.}}: Bare Demo of IEEEtran.cls for Journals}

\maketitle

\begin{abstract}
The compact descriptors for visual search (CDVS) standard from ISO/IEC moving pictures experts group (MPEG) has succeeded in enabling the interoperability for efficient and effective image retrieval by standardizing the bitstream syntax of compact feature descriptors. However, the intensive computation of CDVS encoder unfortunately hinders its widely deployment in industry for large-scale visual search. In this paper, we revisit the merits of low complexity design of CDVS core techniques and present a very fast CDVS encoder by leveraging the massive parallel execution resources of GPU. We elegantly shift the computation-intensive and parallel-friendly modules to the state-of-the-arts GPU platforms, in which the thread block allocation and the memory access are jointly optimized to eliminate performance loss. In addition, those operations with heavy data dependence are allocated to CPU to resolve the extra but non-necessary computation burden for GPU. Furthermore, we have demonstrated the proposed fast CDVS encoder can work well with those convolution neural network approaches which has harmoniously leveraged the advantages of GPU platforms, and yielded significant performance improvements. Comprehensive experimental results over benchmarks are evaluated, which has shown that the fast CDVS encoder using GPU-CPU hybrid computing is promising for scalable visual search.

\end{abstract}



\IEEEpeerreviewmaketitle

\section{Introduction}

\IEEEPARstart{R}{ecently}, there has been an exponential increase in the demand for visual search, which initiates the visual queries to find the images/videos representing the same object or scene. Visual search can facilitate many applications such as product identification, landmark localization, visual odometry, augmented reality, etc. In typical visual-search systems, the users send a query image or its visual feature descriptors to the remote servers\cite{girod2011mobile1}\iffalse \cite{girod2011mobile1,girod2011mobile2}\fi. The images with the same object or scene as that in query image are identified by measuring the visual feature descriptor distance between reference and query images. However, efficient and effective visual search systems are often subject to the constraints of memory footprint, bandwidth and computational cost, low complexity generation and fast transmission of visual queries.
%The well-established architecture performs feature extraction and compression on the client side, such that only compact feature descriptors need to be transmitted to the server side for matching and retrieval. This ensures low bit-rate transmission of visual information and low latency retrieval of visual search.
%increase in the demand for visual search, which initiates the search queries by visual information to find the images/videos representing the same object or scene. Visual search can facilitate many potential applications such as product identification, landmark localization, augmented reality, etc. To realize efficient visual search, new challenges arising from practical applications have emerged. Typically, the visual search is performed at the server side, where there is a database indexing extracted feature descriptors of the reference images \cite{girod2011mobile1,girod2011mobile2}. Due to the stringent constraints on memory, computational power and bandwidth, efficient generation and transmission of the query visual information are critical to improve the user experience. The well-established architecture performs feature extraction and compression on the client side, such that only compact feature descriptors need to be transmitted to the server side for matching and retrieval. This ensures low bit-rate transmission of visual information and low latency retrieval of visual search.

Over the past decade, numerous visual feature descriptors are proposed from the perspectives of high accuracy, low bandwidth and fast extraction. Although the most classical Scale-Invariant Feature Transform (SIFT) descriptor \cite{lowe2004distinctive} has achieved outstanding performance, it imposes severe computational burden and memory cost, especially for mobile visual search or large-scale visual search scenarios.\iffalse, especially for real-time application systems\fi. This led to lots of research work for compact descriptors\iffalse with lower computation and bandwidth requirement\fi. A series of representative visual feature descriptors, e.g., SURF \cite{bay2008speeded}, ORB \cite{Rublee_ORB2011}, BRISK \cite{Leutenegger_2011}, have been proposed. However, most of them approach the goal of reduced computational cost and improved descriptor compactness at the expense of performance loss compared with the original SIFT.

Towards high performance visual search, the Moving Picture Experts Group (MPEG) has published the Compact Descriptors for Visual Search (CDVS) standard in 2015 \cite{duan2016overview}\iffalse\cite{duan2016overview,mpeg_cdvs}\fi. The MPEG-CDVS standard provides the standardized bitstream syntax to enable interoperability for visual search achieving comparable accuracy with much lower bandwidth requirement than SIFT. Herein, two kinds of compressed descriptors, i.e., local and global feature descriptors, are compactly represented at different bit rates to achieve bit-rate scalability (e.g., 512B, 1KB, 2KB, 4KB, 8KB, and 16KB). As such, the stringent bandwidth and accuracy requirements can be well fulfilled.
%by selecting suitable bit rate feature representation and transmission.
%Benefiting from such client-server architecture, the Moving Picture Experts Group (MPEG) has published the Compact Descriptors for Visual Search (CDVS) standard in 2015 \cite{duan2016overview,mpeg_cdvs}. The MPEG-CDVS standard provides the standardized bitstream syntax to enable interoperability for visual search. In CDVS, the handcrafted local and global feature descriptors are compactly represented at different bit rates to achieve bit-rate scalability (e.g., 512B, 1KB, 2KB, 4KB, 8KB, and 16KB). As such, the stringent bandwidth requirements can be satisfied by low bit rate feature representation and transmission.

Besides the bandwidth and accuracy requirements, the encoding efficiency of CDVS descriptors directly determines the visual search latency and affects interactive experience, which is becoming the bottleneck in hindering its wide deployment in industry. Especially, when targeting large-scale video analysis, fast extracting CDVS descriptor from huge amounts of video frames is crucial to support pervasive video analysis applications such as mobile augmented reality, robots, surveillance and media entertainment etc \iffalse\cite{mpeg_cdva_cfp}\fi. Although some algorithms have been proposed to speed up the encoding process in CDVS standard, e.g., image downsampling pre-processing \cite{mpeg_cdvs_standard} and BFLoG \cite{Chen_BFLoG2015}, the efficiency of extracting CDVS descriptors falls far behind the practical requirements of zero-latency or real-time visual search, for example, more than 100 ms per VGA resolution image is incurred on CPU platform.
%an image with both dimensions smaller than 640 pixels.
%In addition to the bandwidth consumption, the extraction speed of CDVS descriptors directly determines the visual search latency and interactive experience. Therefore, high efficiency CDVS implementation and optimization strategies are highly desired in the practical applications of CDVS. With the popularity of large scale video analysis, real time CDVS feature extraction of video frames should also be enabled to support video analysis applications with compact descriptors such as mobile augmented reality, automotive, surveillance, media entertainment, etc \cite{mpeg_cdva_cfp}. Moreover, recently it has been observed that there are complementary effects of CDVS and deep learning based features, and the combination of both approaches can obtain the state-of-the-art performance in image and video analysis \cite{CDVA_Deep,Lou_DCC2017}. As deep learning based features can be efficiently extracted by the graphics processing unit (GPU) \cite{krizhevsky2012imagenet,vasilache2014fast}, how to leverage such infrastructure and achieve high efficiency CDVS feature extraction with GPU is also worth investigated. These emerging requirements all pose new challenges to highly efficient CDVS standardized feature extraction.

Undoubtedly, GPU has achieved great success in high throughput image and video processing due to its parallel-processing capability \cite{garland2008parallel}. Especially for the state-of-the-arts convolution neural network (CNN) approaches, GPU has become the crucial computation platform. Therefore, how to leverage GPU to significantly speed up CDVS encoder, and explore the harmonious operation and complementary effects of (handcrafted) CDVS compact descriptors and deep learning based features over GPU platform is becoming a promising and practically useful topic. In this paper, we first revisit the CDVS technique contributions in reducing computational cost. Then, we present the fast MPEG-CDVS encoder. The main contributions of this paper are three-fold:
%By thorough analysis on the calculations of CDVS, we shift the most time-consume and parallel-friendly modules to the GPU platform and perform the data-dependent calculations on CPU to efficiently utilize the two computation architectures. The main contributions of this paper are as follows:
% Recently, the popularity of GPU \cite{garland2008parallel}, which provides the parallel-processing capabilities for for highly data parallel applications, has been creating a strong demand in GPU based CDVS implementation. In order to further push CDVS towards industry applications, this paper performs a thorough analysis on the complexity of CDVS and provides a practical strategy for GPU based CDVS implementation. The main contributions of this paper are as follows:
\begin{enumerate}
\item We revisited significant contributions of MPEG-CDVS from the perspective of reducing the computational cost of CDVS encoders, and its merits of accommodating parallel implementation over GPU-CPU hybrid computing platforms. With the challenges of big image/video data analysis, the exploration of high throughout computing of standard compliant low complexity CDVS descriptor (or other handcrafted features) via hybrid platforms are expected to facilitate the deployment of scalable and interoperable visual search applications.

\item We proposed a very fast CDVS encoder, which elegantly shifts the computational intensive operations to GPU platform. By leveraging the high parallel processing capability of GPU and the strength of parallel operations in CDVS, the fast CDVS encoder has achieved up to $30\times$ speedup over CPU platform without noticeable performance loss. To the best of our knowledge, this is the first and the fastest CDVS standard compliant encoder over GPU platforms.

% Leveraging thousands of parallel processors, GPU has achieved great success in accelerating computation-intensive algorithms by parallelizing them on massive processors and performing them simultaneously. This paper proposed a well designed fast CDVS encoder using GPU-CPU hybrid computing to speed up CDVS by shifting the computation-intensive operations to GPU platform. Due to the high parallel capability of GPU and the parallel-friendly computations of CDVS, the proposed fast CDVS achieves significant speedup compared with that on CPU platform without performance deterioration.

% An efficient CDVS implementation using GPU-CPU hybrid computing is proposed to solve the speedup problem for CDVS. The CDVS is significantly accelerated by well designing thread block allocation and optimized the memory access to make use of the GPU computational resources sufficiently and efficiently.
%GPU has a massively parallel architecture consisting of thousands of smaller, more efficient cores designed for handling multiple tasks simultaneously
%An efficient CDVS implementation using GPU-CPU hybrid computing is proposed to solve the speedup problem for CDVS. The CDVS is significantly accelerated by well designing thread block allocation and optimized the memory access to make use of the GPU computational resources sufficiently and efficiently.

\item Furthermore, we have studied the significant performance improvement by combining CDVS descriptors and CNNs features over benchmarks, with 0.0305 and 0.174 mAP gains over CDVS or CNNs, respectively. In particular, we propose the marriage of higher computational efficiency of CDVS (3.27 ms CDVS vs 144 ms CNNs for a $640\times480$ image) and promising search performance of CNNs towards scalable visual search framework, in which their complementary effects in terms of efficiency and performance have been well demonstrated.

\end{enumerate}

The remainder of this paper is organized as follows. Section II reviews the related works. Section III revisits the techniques to speed up the CDVS encoding process. Section IV presents the fast CDVS encoder using GPU-CPU hybrid computing. Extensive experimental results and discussions are reported in Section V, and finally we conclude this paper in Section VI.

\section{Related Work}

\subsection{Introduction of GPU Architecture}
\begin{figure}[t]
\centering
%\begin{minipage}[b]{0.435\textwidth}
\includegraphics[width=8.8cm]{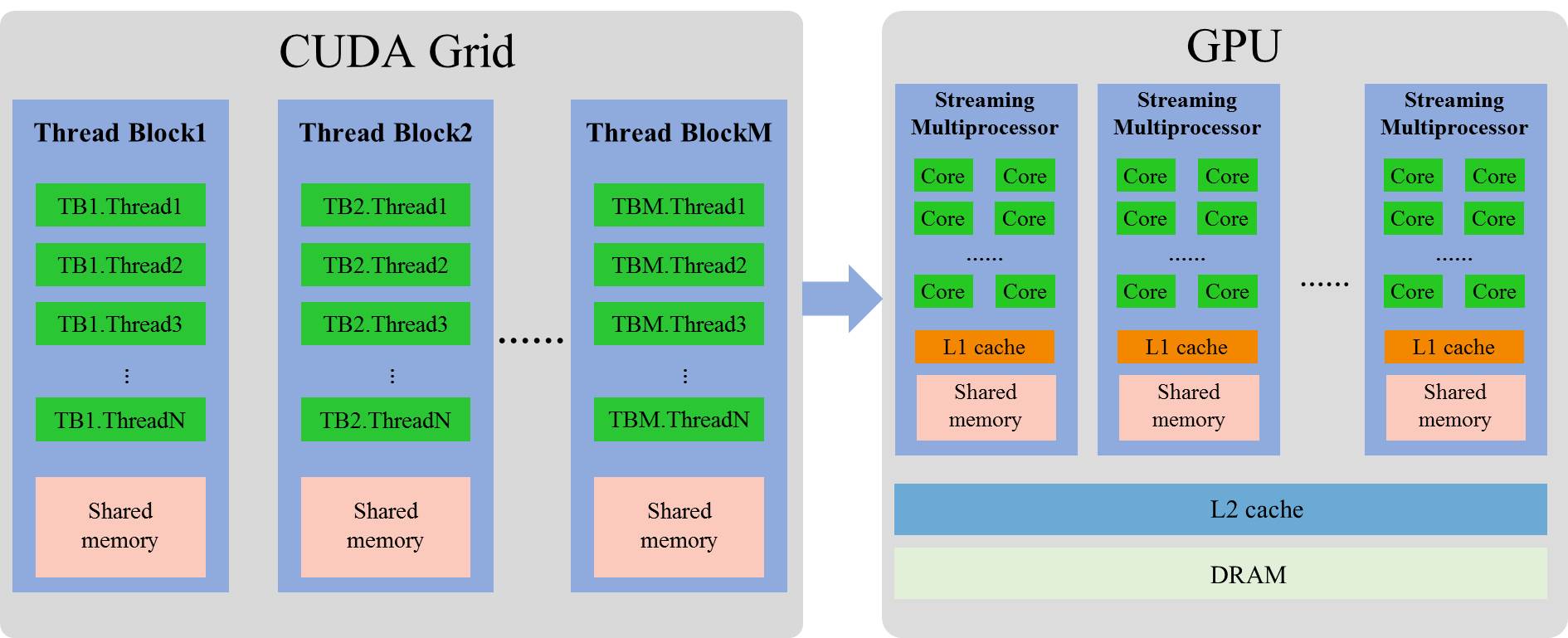}
%\subfigure[]{\label{fig:ratemodel_b}\includegraphics[width=7.0cm]{DistortionReduction.jpg} }
%\end{minipage}
\caption{Thread Batching and GPU architecture. The computations are executed using a batch of threads organized as a grid of thread blocks, which are allocated to streaming multiprocessors on GPU.\iffalse Thread Batching, the host execute an kernel using a batch of threads organized as a grid of thread blocks\fi }
\label{fig:threadbatching}
\end{figure}

Different from CPU, consisting of a few cores optimized for sequential processing, GPU exhibits massively parallel architecture consisting of thousands of smaller, but more efficient cores designed for handling multiple tasks simultaneously by launching with Single Instruction Multiple Threads (SIMT) in which a set of atom operations are applied to process huge amounts of pixels in parallel. There exist a variety of parallel computing platforms and application programming interface models created in recent years, e.g., CUDA, Directcompute and OpenCL, which have significantly strengthened the parallel-processing capabilities of GPUs towards general-purpose computing. Herein, CUDA is the most widely used parallel programming framework developed by NVIDIA. It partitions workloads into thread blocks (TB), each of which is a batch of threads that can cooperate together by efficiently sharing data through some fast shared memory and synchronizing their execution to coordinate memory accesses. Furthermore, thread blocks of same dimensionality and size that execute the same kernel can be batched together into a grid of blocks, so that the total number of threads that can be launched in a single kernel invocation is much larger as illustrated in Fig.~\ref{fig:threadbatching} \cite{clara2008nvidia}. The TBs are allocated to streaming multiprocessors (SM) to be executed simultaneously using GPU cores. In addition, an important feature for the shared memory is that the memory access operation can be performed simultaneously when the adjacent threads in the same TB access the adjacent shared memory units, which is known as \textit{global memory coalescing}. Therefore, by optimizing the TB allocation and memory access, the speedup of calculations on GPU can be further improved.

\begin{figure*}[t]
\centering
%\begin{minipage}[b]{0.435\textwidth}
\includegraphics[width=16.0cm]{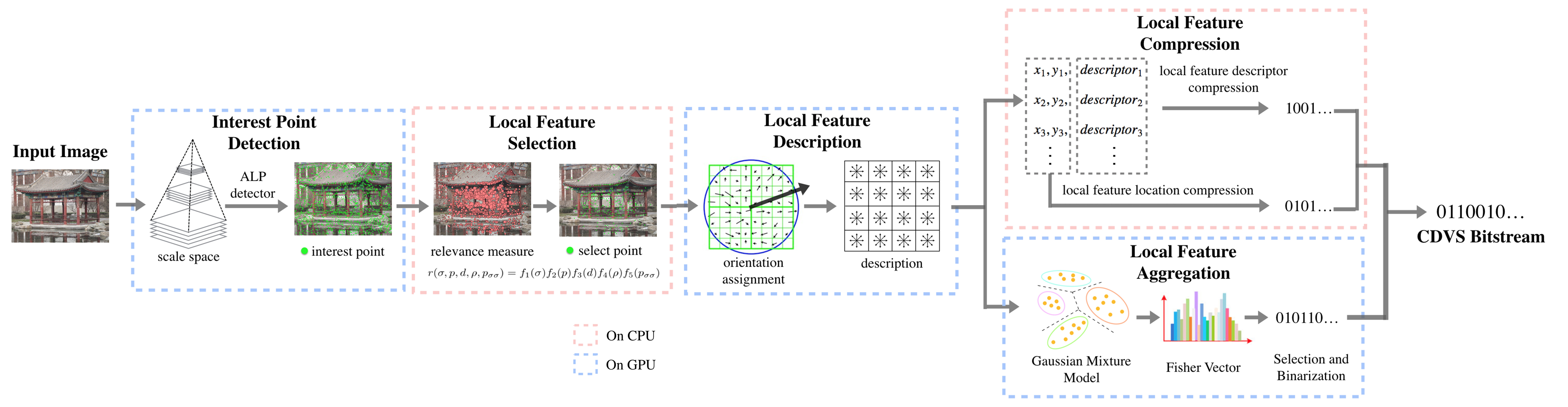}
%\subfigure[]{\label{fig:ratemodel_b}\includegraphics[width=7.0cm]{DistortionReduction.jpg} }
%\end{minipage}
\caption{The normative blocks of the CDVS standard implemented on GPU-CPU hybrid computing architecture. }
\label{fig:NormBlockCDVS}
\end{figure*}

\subsection{Review of GPU based feature extraction}

Based on the outstanding parallel performance of GPU, there has been a fast growing interest in applying GPU to speed up visual feature descriptor construction. Especially, tremendous algorithms have been proposed for GPU based SIFT implementation \cite{heymann2007sift,wu2007siftgpu,rister2013fast,lee2016complexity,cudasift,wang2013workload,patlolla2015gpu}, as SIFT descriptors require high computational cost and huge demand of memory. In \cite{heymann2007sift}, an efficient GPU implementation of SIFT was presented based on the vector operations of GPU by texture packing, and 20 fps with the QuadroxFX 3400 GPU was realized. In \cite{wu2007siftgpu}, an open source GPU/CPU mix implementation for SIFT was provided and achieved 27.1 fps with CUDA on 8800GTX GPU. In \cite{cudasift}, with GTX 1060 GPU the CUDA-SIFT implementation consumes 2.7 ms for the images with resolution \(1280\times960\) and 3.8 ms for the resolution \(1920\times1080\). Wang \textit{et al.} further analyzed the workload of SIFT in \cite{wang2013workload} and proposed to distribute the feature extraction tasks to CPU and GPU, such that a speed of 10 fps for a $320\times 256$ image and 41\% energy consumption reduction can be achieved. Besides SIFT, the speeded-up robust feature (SURF) \cite{bay2008speeded, cornelis2008fast} and fisher vectors (FV) \cite{ma2016gpu} were also explored in implementing using GPU platform, and around an order of magnitude speedup was achieved compared to CPU based implementation.

Besides hand-crafted features, recently convolutional neural network based features have achieved promising performance in various computer vision tasks such as image classification \cite{krizhevsky2012imagenet} and retrieval \cite{zheng2016sift}\iffalse face recognition \cite{sun2014deep}, image classification \cite{krizhevsky2012imagenet}, retrieval \cite{zheng2016sift,Lou_DCC2017}, etc\fi. CNN requires extremely fast parallel feature extraction on GPU. In \cite{krizhevsky2012imagenet}, a highly optimized GPU implementation of CNN is made publicly available for training networks. A number of CNN softwares based on CUDA using NVIDIA GPUs have also been developed, such as Caffe \cite{jia2014caffe} and TensorFlow \cite{abadi2016tensorflow}. \iffalse Recently, Vasilache \textit{et al.} introduced the fast fourier transform convolution implementation based on NVIDIA's cuFFT library \cite{vasilache2014fast}, which is faster than than NVIDIA's cuDNN implementation for the common convolutional layers.\fi

However, although there are many visual feature descriptors implemented on GPU, they are inferior in fulfilling several practical but crucial requirements, e.g., low bandwidth cost, high performance, good compactness, and the excessive GPU resource consumption. Therefore, the very fast and standard compliant CDVS encoder over GPU platforms can elegantly contribute to the state-of-the-art large-scale visual search.

% In addition to the hand-crafted features, due to the powerful computational ability of GPU, recently deep learning based features have achieved promising performance in various computer vision tasks such as face recognition \cite{sun2014deep}, image classification \cite{krizhevsky2012imagenet}, retrieval \cite{zheng2016sift,Lou_DCC2017}, etc. The convolutional neural network can perform an extremely fast parallel feature extraction due to its design architecture in nature. In \cite{krizhevsky2012imagenet}, a highly optimized GPU implementation of CNN is made publicly for training newtworks for image classification. A number of deep learning software based on CUDA using NVIDIA GPUs have also been developed, such as Caffe \cite{jia2014caffe}, TensorFlow \cite{abadi2016tensorflow}, etc. Recently, Vasilache \textit{et al.} introduced the fast fourier transform convolution implementations based on NVIDIA's cuFFT library \cite{vasilache2014fast}, which is faster than than NVIDIA's cuDNN implementation for the common convolutional layers.

%\hfill December 27, 2012

\section{CDVS Revisit for Speedup}

Targeting high accuracy and low bandwidth, CDVS has achieved significant success. Although several technique proposals \iffalse \cite{ISO2013_MPEGCDVS28076,ISO2013_MPEGCDVS28090,ISO2013_MPEGCDVS30241,ISO2014_MPEGCDVS33159,ISO2013_MPEGCDVS31399,ISO2013_MPEGCDVS30256,Chen_BFLoG2015}\fi \cite{ISO2013_MPEGCDVS28076,ISO2013_MPEGCDVS28090,ISO2014_MPEGCDVS33159,ISO2013_MPEGCDVS31399,ISO2013_MPEGCDVS30256,Chen_BFLoG2015} were proposed to speed up the encoding process of CDVS, it cannot fulfill real-time requirement on CPU platform. To reduce the computational complexity, CDVS first downsamples the input images into low resolution with the longer side less than 640. However, CDVS extraction still needs more than 100 ms for one image \cite{Chen_BFLoG2015}. %In this section, we revisit these contributions for CDVS acceleration to explore the room of further speedup.

\begin{figure}[t]
\centering
%\begin{minipage}[b]{0.435\textwidth}
\includegraphics[width=7.0cm]{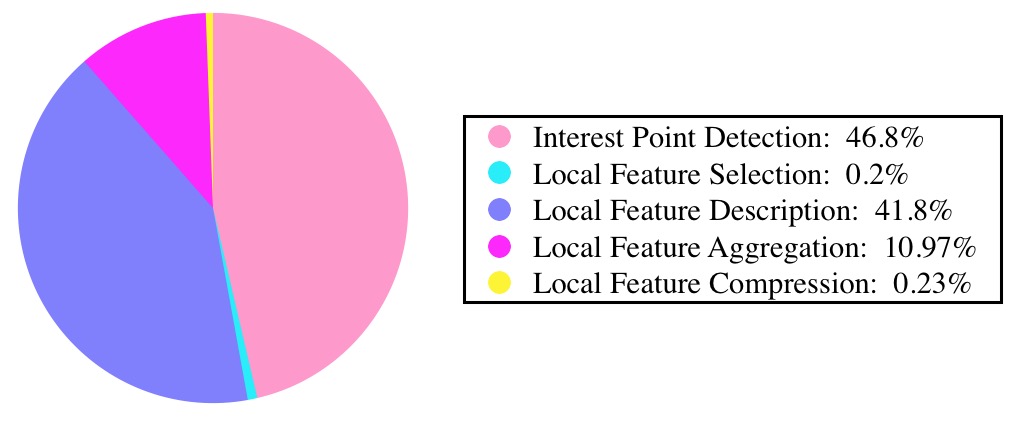}
%\subfigure[]{\label{fig:ratemodel_b}\includegraphics[width=7.0cm]{DistortionReduction.jpg} }
%\end{minipage}
\caption{The running time consumption for different modules of CDVS, tested on Linux PC with Intel(R) Xeon(R) CPU E5-2650 v2@2.60GHz. }
\label{fig:TimeDistribution}
\end{figure}

As illustrated in Fig.\ref{fig:NormBlockCDVS}, the CDVS encoder can be divided into five major modules, i.e., interest point detection, local feature selection, description, compression and aggregation. To analyze the computation cost of different modules, we test the running time of each module using TM14.0 on 1000 images with resolution $640\times 480$. Fig.\ref{fig:TimeDistribution} shows the percentage of average running time for different modules. From the results we can see that interest point detection, local feature description and aggregation take up most of the running time, more than 99.5\%. Therefore, the optimization strategy of CDVS encoders are centralized in the three modules.

\subsection{Interest Point Detection}

The interest point detection consists of two stages, i.e., scale-space construction and extremum detection. The CDVS constructs the scale space as an image pyramid which is generated by filtering the input image via a series of 2D separable Gaussian filters at increasing scale factors. The interest points are regarded as the scale-space extrema of the normalized derivatives of each scale in an image pyramid, which is generated by applying Laplacian filter to each scale image. Therefore, there are multiple convolution operations for input image $I$ with different scale factors as follows,
\begin{equation}
L_{k} = I * g_{k} * f,
\label{Eqn:LoG}
\end{equation}
where $g_{k}$ and $f$ are the Gaussian and Laplacian kernels, respectively. Obviously, it is a computation intensive process.

To speed up the process of LoG filtering, the Block based Frequency Domain Laplacian of Gaussian (BFLoG) filtering \iffalse\cite{ISO2013_MPEGCDVS28076,ISO2013_MPEGCDVS28090,ISO2013_MPEGCDVS30241,ISO2014_MPEGCDVS33159,Chen_BFLoG2015}\fi \cite{ISO2013_MPEGCDVS28076,ISO2013_MPEGCDVS28090,ISO2014_MPEGCDVS33159,Chen_BFLoG2015} is proposed instead of that in spatial domain. The original input image is first decomposed into overlapped blocks, which are further transformed into frequency domain using Discrete Fourier Transform (DFT). Then, the spatial domain convolution operation can be equivalently implemented by element-wise product between the frequency image matrix and frequency filter kernel matrix\iffalse, as illustrated in Fig.\ref{fig:BFLoG}\fi. To remove the FFTs for filter kernels, BFLoG adopts the fixed block size to pre-compute convolution kernels in DFT domain, and the pre-computation also reduces the memory cost. Due to Fast Fourier Transform (FFT) \cite{carr1999option}, the computational complexity of convolution in spatial domain $\mathcal{O}(M^2N^2)$ becomes $\mathcal{O}(4N^2logN^2-11N^2+16N)$, where $M$ and $N$ are the size of square filter and square block. By optimizing the block size and overlap size, the BFLoG achieves about 47\% filtering time reduction with ignorable performance variations.\iffalse the BFLoG achieves significant running time reduction for CDVS with ignorable variations in visual search performance, achieving about 47\% running time reduction compared with spatial filtering\fi \iffalse as show in Table \ref{tab:BFLoG_TimeReduction}.\fi

Although the block-level interest point detection are independent of each other, the pixels of each block are dependent in FFT calculation, which makes it difficult to be implemented on pixel-level parallelism. To reduce the boundary effects, BFLoG utilizes $R$ overlapped pixels for each $N\times N$ block, which leads to extra computational burdens. Empirically, the block size $N=128$ and overlapped size $R=16$ are optimal for performance and efficiency. However, for an $640\times 480$ image, only 35 blocks can be executed in parallel, which is much fewer than the number of GPU cores.

%\begin{figure}[t]
%\centering
%%\begin{minipage}[b]{0.435\textwidth}
%\includegraphics[width=7.0cm]{BFLoG.jpg}
%%\subfigure[]{\label{fig:ratemodel_b}\includegraphics[width=7.0cm]{DistortionReduction.jpg} }
%%\end{minipage}
%\caption{The image pyramid construction using BFLoG. }
%\label{fig:BFLoG}
%\end{figure}

%\begin{table}[htbp]
%  \centering
%  \renewcommand{\arraystretch}{1.1}
%  \caption{Comparison of running time for scale-space construction using BFLoG and spatial domain DoG (Uint: ms), tested on Windows PC with INTEL core CPU I5 3470@3.2 GHz \cite{Chen_BFLoG2015}.}
%    \begin{tabular}{c|c|c|c|c|c}
%     \hline
%    %\toprule
%    \multicolumn{2}{c|}{Octave} & \multicolumn{3}{c|}{BFLoG} & \multirow{2}{*}{Spatial filtering} \\  \cline{1-5}
%    W    & H   & FFTs   & Filtering   & Total       \\\hline
%    640  & 480   & 113.86  & 2.29    & 116.16 & 259.28\\ \hline
%    320  & 240   & 39.04     & 0.79    & 39.83 & 55.6 \\ \hline
%    160  & 120   & 13.01     & 0.26    & 13.28 & 13.9\\ \hline
%    80   & 60   &  3.25     & 0.007   & 3.32   & 3.48\\ \hline
%    40   & 30   &  3.25     & 0.007   & 3.32   & 0.87\\ \hline
%\multicolumn{2}{c|}{Total} & 172.42    &  3.47     &  175.9    & 333.12 \\ \hline
%    %\bottomrule
%    \end{tabular}%
%  \label{tab:BFLoG_TimeReduction}%
%\end{table}%

\begin{figure}[t]
\centering
\begin{minipage}[b]{0.4\textwidth}
\subfigure[]{\label{fig:BMLoG_a}\includegraphics[width=7.0cm]{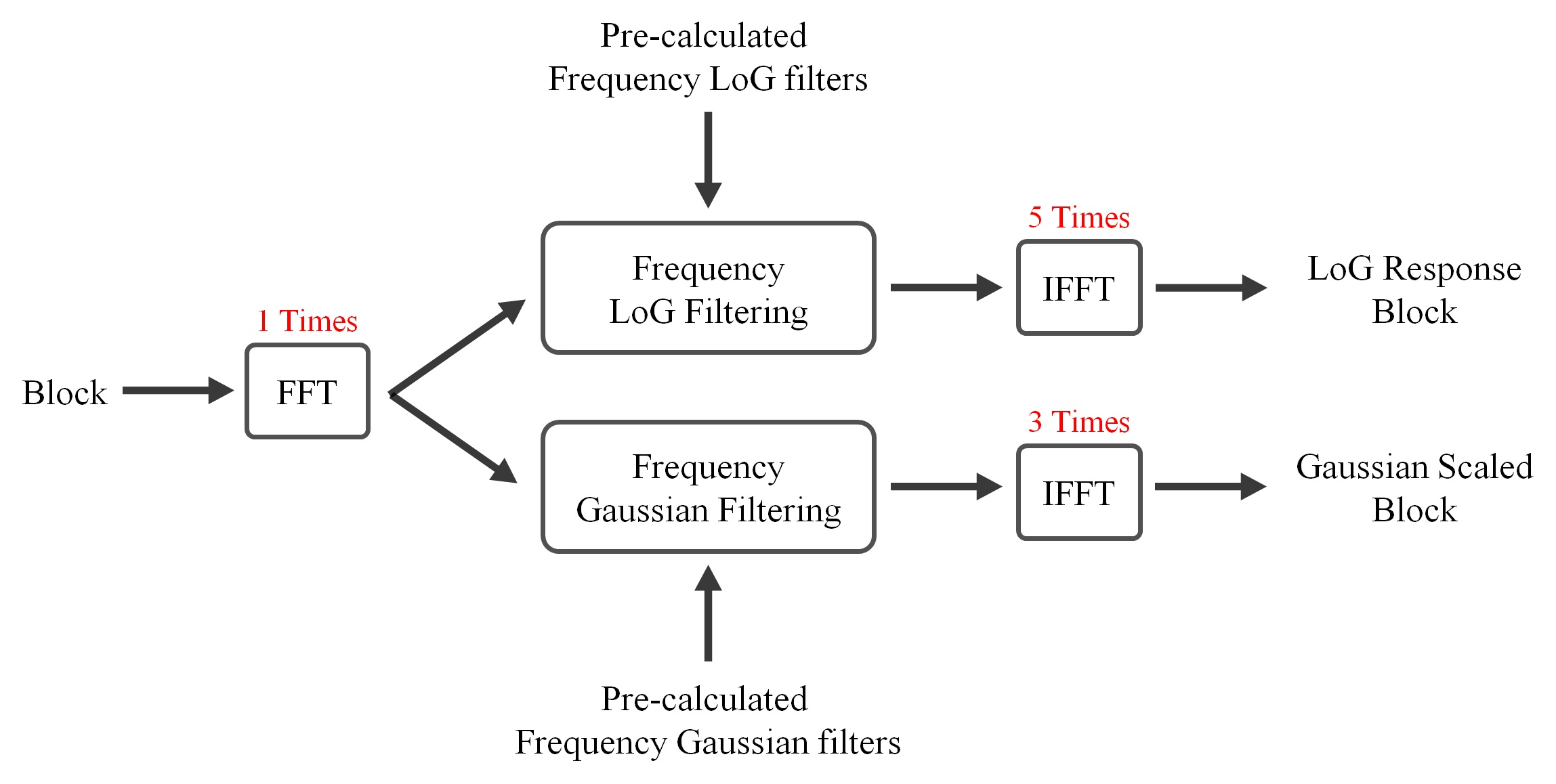} }
\end{minipage}
\begin{minipage}[b]{0.4\textwidth}
\subfigure[]{\label{fig:BMLoG_b}\includegraphics[width=7.0cm]{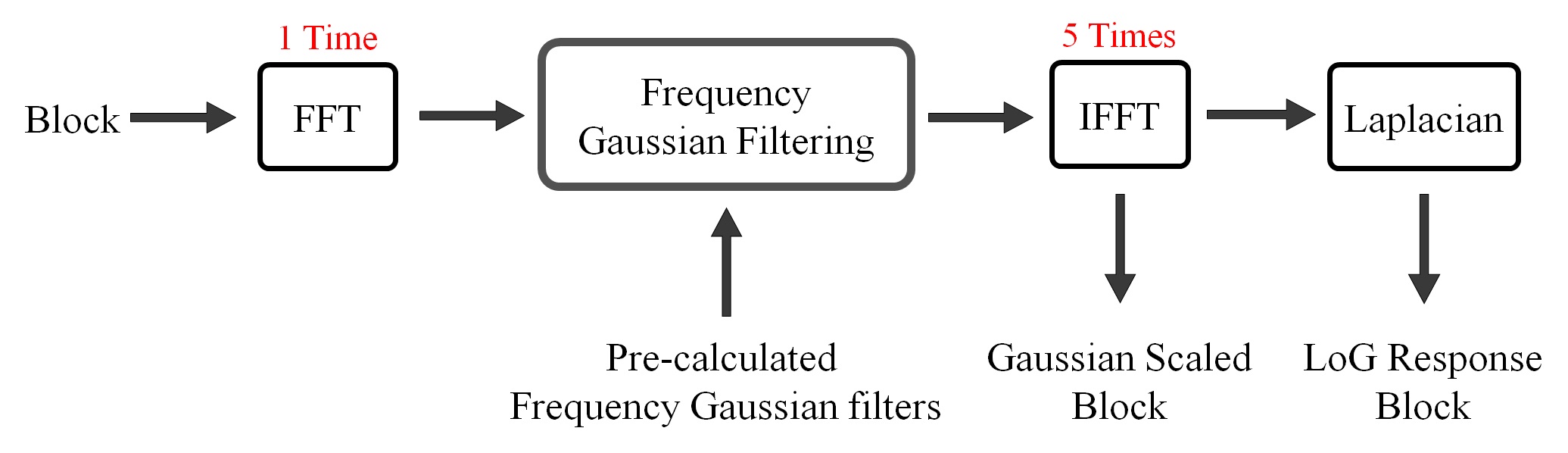} }
\end{minipage}
\caption{The filtering approach for BFLoG and BMLoG, (a) BFLoG, (b) BMLoG. }
\label{fig:BMLoG}
\end{figure}

For the subsequent extremum detection, \iffalse orientation assignment and descriptor calculation,\fi BFLoG has to recompose the (LoG and Gaussian) filtered representations, which leads to extra IFFT operations and memory increase. There are 8 inverse IFFTs for each block with 5 scales as illustrated in Fig.\ref{fig:BMLoG_a}. Considering the Laplacian filter kernel with fixed and small size for all the scale images, the computational cost of Laplacian convolution is much lower than that of Gaussian convolution. Therefore, Duan \textit{et al.} proposed the mixture domain LoG filtering approach (BMLoG) to further reduce the computation cost of BFLoG by applying Gaussian convolution in frequency domain and Laplacian convolution in spatial domain as shown in Fig.\ref{fig:BMLoG_b}. As a result, the filtering process has 1 FFT and 5 IFFT operations plus $5\times 7$ adds/subtractions of each pixel for the Laplacian convolution, which achieves about 20\% filtering time reduction \cite{ISO2013_MPEGCDVS31399}.

After the LoG filtering, interest points are detected by identifying the local extrema, which needs to compare each sample point with its 8 neighbors in the current scale image and 18 neighbors in the above and below scales. CDVS adopts an alternative extrema detection algorithm with low computational complexity, which is a low-degree polynomial (ALP) approach \cite{ISO2013_MPEGCDVS30256}. By assuming that LoG kernel can be approximated by linear combinations of LoG kernels at different coordinates and scales, ALP approximates the LoG scale space by a third-degree polynomial of the scale $\sigma_k$ for each sample point $(x,y)$,
\begin{equation}
%p(x,y,\sigma) = \alpha_3(x,y)\sigma^3 + \alpha_2(x,y)\sigma^2 + \alpha_1(x,y)\sigma + \alpha_0(x,y)
p(x,y,\sigma_k) = \sum_{i=0}^3\alpha_i(x,y)\sigma_k^i,
\label{Eqn_ALP}
\end{equation}
where the polynomial coefficients are functions of the image coordinates $(x,y)$,
\begin{equation}
%\begin{split}
\alpha_i = \sum_{k=0}^{K-1}\beta_{k,i}L_k(x,y),~~ i= 0, 1, 2, 3.
%\alpha_3(x,y) =  \sum_{k=0}^{K-1}a_kL_k(x,y), \alpha_2(x,y) =  \sum_{k=0}^{K-1}b_kL_k(x,y), \\
%\alpha_1(x,y) =  \sum_{k=0}^{K-1}c_kL_k(x,y), \alpha_0(x,y) =  \sum_{k=0}^{K-1}d_kL_k(x,y),
\label{Eqn_ALP_Coef}
%\end{split}
\end{equation}
The parameters, $\{\beta_{k,i}\}$, correspond to the $K$ predefined scales $\sigma_k$ and $\{L_k|k=0,\cdots,K\}$ are the LoG filtered images in one octave. ALP first finds the scales of the extrema via the derivatives with respect to $\sigma_k$ of the polynomial in Eqn.(\ref{Eqn_ALP}), and then it compares the point with its 8 neighbors. Compared with the previous methods, ALP is more efficient by reducing the 18 comparisons for each sample point to 8 comparisons. %In addition, the above operations can be done independently for all pixels and are well suited for highly parallel architectures like the GPU.

Although these fast algorithms significantly reduce both the computational and memory costs, the computational burden of interest point detection is still too high for CPU, and the interest point detection is still the most time-consuming module. Fortunately, the calculations of convolution and ALP are suitable for pixel-level parallelism. These computations can well fit into highly parallel architectures like GPU.

%where $S$ is a scale parameter, which has two options, \textit{i.e.},
%\begin{equation}
%S =  2B(x,y) + \frac{-2B(x,y)\pm(4B^2(x,y)-12A(x,y)C(x,y))}{6A(x,y)}.
%\label{Eqn_localresponse_comp}
%\end{equation}
% The two options are used to determine current response is the minimum and maximum values, respectively. These computations is repeatedly performed in every octave, and for neighboring octaves, the duplicate elimination should be carried out on all the candidate interest points. This is really a time-consuming module because there are lots of comparison and multiplication operations, and its computational complexity is around $\mathcal{O}(c_1M^2WH+c_2WH)$, where $c_1$ and $c_2$ are two constants related with the number of scale images and octaves. Fortunately, the data involved into these operations are independent, which makes the above operations can be well implemented in parallel.

\subsection{Local Feature selection and Description}

\begin{figure}[t]
\centering
\includegraphics[width=5.5cm]{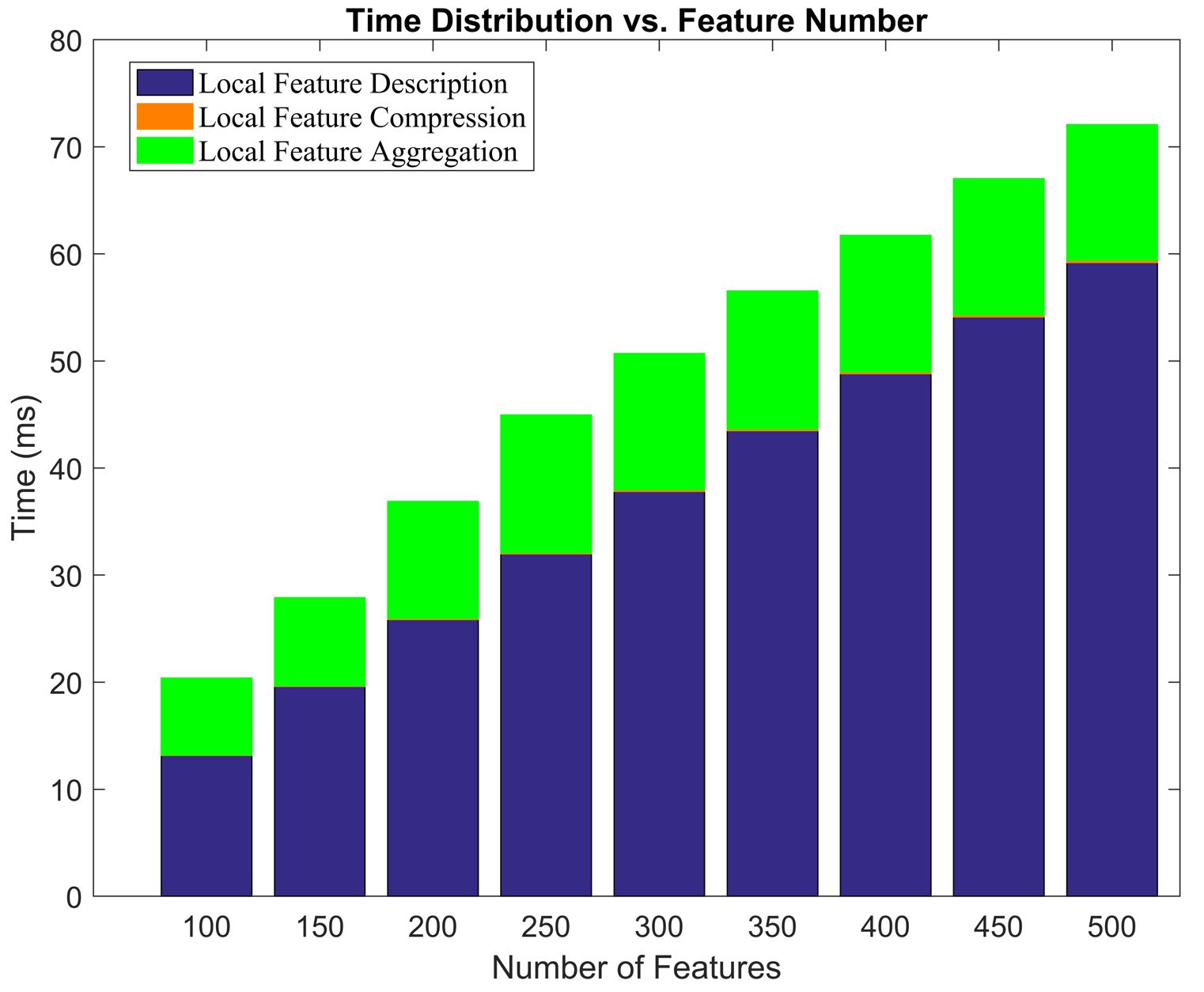}
\caption{The time consumption of CDVS modules without local feature selection when extracting different number of local features. The test is performed on Linux PC with Intel(R) Xeon(R) CPU E5-2650 v2@2.60GHz for 1000 images with resolution $640 \times 480$.}
\label{fig:featureselectionTimeNumber}
\end{figure}

\begin{figure}[t]

  \begin{minipage}[b]{0.235\textwidth}
  \centering
\subfigure[\emph{Interest points in query image}]{\label{fig:selection_a}\includegraphics[width=4.0cm]{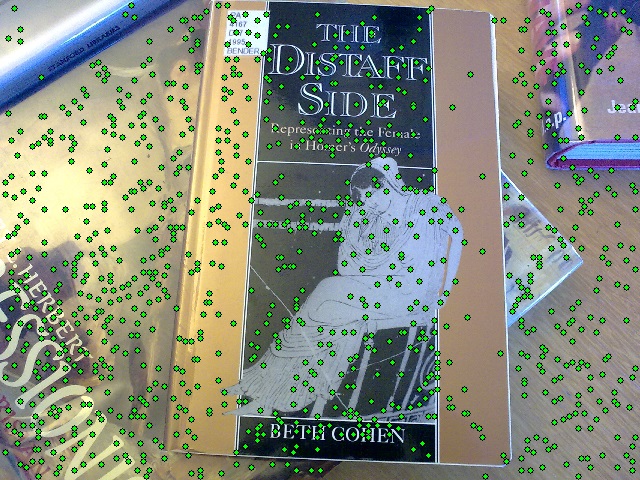} }
  \end{minipage}
  \hspace{0.1cm}
  \begin{minipage}[b]{0.235\textwidth}
  \centering
  \subfigure[\emph{Relevance of the interest points}]{\label{fig:selection_b}\includegraphics[width=4.0cm]{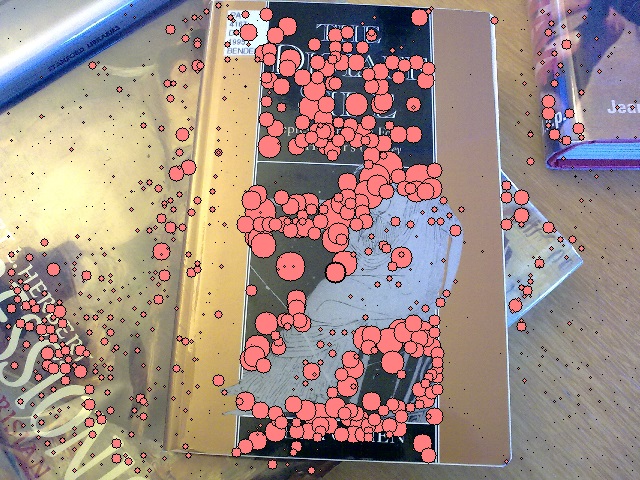}  }
  \end{minipage}

  \caption{Comparison of ALP interest point distribution and the selected interest point distribution using the ranked values in Eqn.(\ref{Eqn_relevance}), (a) ALP interest points, (b) relevance of the interest points, where the circle size of interest points are proportional to the relevance measure.}
\label{fig:featureselection}
\end{figure}

Since there are usually hundreds or thousands of interest points in an image as illustrated in Fig.~\ref{fig:featureselection}, it brings difficulties in compact representation and low computational cost when processing all features. Fig.~\ref{fig:featureselectionTimeNumber} shows the relationship between computational cost and feature number. Selecting a subset of good features may save considerable computation time for the subsequent local feature description, compression and aggregation. \iffalse The computational cost of local feature description and aggregation increase obviously along with the feature number. \fi Therefore, the core module of local feature selection plays an important role in reducing the computational cost.%, which selects a subset of interest points to describe and compress while keeping high accuracy for visual search. Then, it can save considerable computation time for the subsequent local feature description, compression and aggregation due to the limited feature number.

During the development of CDVS, lots of methods \cite{ISO2012_MPEGCDVS23822,ISO2012_MPEGCDVS23929,francini2013selection} are proposed to describe and compress a subset of essential interest points while maintaining and even improving search accuracy. The basic rationale is based on the statistical characteristics to select the interest points with high probability being positive matching ones and remove the noise points. Samsung Electronic proposed to extract features in visual attention regions \cite{ISO2012_MPEGCDVS23822} based on the assumption that more relevant descriptors are located in salient regions for human visual system. Thus, computation overhead can be significantly reduced by applying feature extraction to regions of interest (ROI). However, this method leads to performance loss due to the inconsistency between ROIs and the distribution of true matching points. Moreover, it is difficult to define accurate ROIs at low computational cost.
%\begin{figure}[t]
%
%  \begin{minipage}[b]{0.235\textwidth}
%  \centering
%\subfigure[\emph{Interest points in query image}]{\label{fig:selection_a}\includegraphics[width=4.2cm]{1query.jpg} }
%  \end{minipage}
%  \hspace{0.1cm}
%  \begin{minipage}[b]{0.235\textwidth}
%  \centering
%  \subfigure[\emph{Interest points in reference image}]{\label{fig:selection_b}\includegraphics[width=2.10cm]{2reference.jpg} }
%  \end{minipage}
%    \begin{minipage}[b]{0.435\textwidth}
%  \centering
%  \subfigure[\emph{Matching points}]{\label{fig:selection_c}\includegraphics[width=7.2cm]{s.jpg} }
%  \end{minipage}
%  \begin{minipage}[b]{0.235\textwidth}
%  \centering
%  \subfigure[\emph{Relevance of the interest points}]{\label{fig:selection_d}\includegraphics[width=4.2cm]{4relevance.jpg} }
%  \end{minipage}
%  \begin{minipage}[b]{0.235\textwidth}
%  \centering
%  \subfigure[\emph{Matched points in query image}]{\label{fig:selection_e}\includegraphics[width=4.2cm]{5featureselection.jpg} }
%  \end{minipage}
%
%  \caption{Comparison of ALP interest point distribution and the selected interest point distribution using the ranked values in Eqn.(\ref{Eqn_relevance}), (a) ALP interest points in query image, (b) ALP interest points in reference image, (c) matching points between query and reference images, (d) relevance of the interest points, (e) the matched point distribution, the red ones are in feature selection set and the green ones are not in the feature selection set, 88\% matched points are in the feature selection set.}
%\label{fig:featureselection}
%\end{figure}

Finally, CDVS adopts a relevance measure to evaluate feature significance in image matching and retrieval performance based on five statistical characteristics of interest points, including the scale $\sigma$ of the interest point, the scale-normalized LoG response value $p$ obtained with $p(x,y,\sigma)$ as in Eqn.(\ref{Eqn_ALP}), the distance $d$ from the interest point to the image center, the ratio $\rho$ of the squared trace of Hessian to the determinant of the Hessian, and the second derivative $p_{\sigma\sigma}$ of the scale space function with regard to $\sigma$, i.e., $\frac{\partial^{2}p(x,y,\sigma)}{\partial^2\sigma}$. The relevance measure indicates the priori probability of a query feature correctly matching a feature of database images. Given a characteristic parameter by symbol $y$ lying within region $B$, the conditional probability for correct matching is
\begin{equation}
f(c=1|y\in B) = \frac{P(y\in B, c=1 )}{P(y\in B)}.
\label{Eqn_conditionalprob}
\end{equation}
CDVS attempted to learn these conditional distributions over training pairwise feature matching from a large database of matching image pairs\cite{francini2013selection}, and stored the quantized characteristic parameters and their corresponding function values of conditional distributions in normative tables, which may minimize the computational cost by look-up tables.

By assuming independency of the characteristic parameters, the relevant score for each point is obtained by multiplying these conditional probabilities:
\begin{equation}
r(\sigma,p,d,\rho,p_{\sigma\sigma}) = f_1(\sigma)f_2(p)f_3(d)f_4(\rho)f_5(p_{\sigma\sigma}).
\label{Eqn_relevance}
\end{equation}
Therefore, a subset of $N$ interest points will be selected by ranking relevance measure $r$, and the feature description is only performed on the selected points. Fig.\ref{fig:featureselection} shows the distribution comparison before and after selecting salient interest points. \iffalse Although there are lots of interest points in images, only a few of them are useful in matching.\fi The circle size of interest points in Fig.\ref{fig:selection_b} are proportional to the relevance value. We can see that the selected relevant interest points by Eqn.(\ref{Eqn_relevance}) basically fall into the salient regions, in which feature selection can significantly reduce the computational cost without performance degeneration. To balance the computational cost and search accuracy, CDVS empirically selects around 300 local feature descriptors to represent an image, which saves more than 30\% computational cost (seeing Fig.\ref{fig:featureselectionTimeNumber}).
%Therefore, a subset of $N$ interest points will be selected on the basis of the ranked relevance measure $r$, and then the feature description is only performed on the selected points. Fig.\ref{fig:featureselection} shows the comparison of the ALP interest point distribution and the selected interest point distribution. Although there are lots of interest points in query and reference images, only a few of them are useful in matching. The circle size of interest points in Fig.\ref{fig:selection_c} are proportional to the relevance measure, and the points in Fig.\ref{fig:selection_e} are the matched points in query image. \iffalse In addition, Fig.~\ref{fig:selection_c} illustrates an pair-wise matching example with geometric consistency check, and the left image is the same with that of Fig.~\ref{fig:selection_a} denoted as query image and the right is an reference one.\fi We can see that selected relevant interest points according the score in Eqn.(\ref{Eqn_relevance}) are quite approximate to the matched points in practice, which makes the local feature selection can directly reduce the computational complexity of CDVS without degeneration of visual search performance.

%In view of the performance, the CDVS adopts the SIFT descriptor to represent the visual information, which needs to calculate the dominant orientations and orientation histogram for each selected interest point.

\subsection{Local Feature Compression}

The uncompressed SIFT descriptors are difficult to use in practice for two reasons, 1) size limitation: it needs 1024 bits for each descriptor by representing each dimension with 1 byte; 2) speed limitation: the computational cost for byte-vector distance is too high over large-scale databases. Therefore, the local feature compression and fast matching in compressed domain are necessary.

%The uncompressed SIFT descriptors are difficult to be applied in visual search applications for the two reasons, 1) bandwidth limitation: it needs 1024 bits for each descriptor by representing each dimensional with 1 byte; 2) speed limitation: the computational complexity for byte-vector distance is too high for visual search over large-scale database. Therefore, high efficiency compression for local feature descriptor is necessary for compacting represent features by reducing the redundancy inside local feature descriptor while well keeping their discriminative power.

During CDVS development, two kinds of compression algorithms based on vector quantization and transform are widely discussed and both of them achieve significant improvement on compression performance and computational efficiency. In early stage of CDVS, the Test Model under Consideration (TMuC) for CDVS \cite{ISO2012_MPEGCDVS12367} employed tree structured vector quantization (TSVQ) \cite{ISO2012_MPEGCDVS24737} and product quantization (PQ) \cite{jegou2011product} to make compact descriptors. However, these methods need to store huge codebooks, thereby leading to heavy computational burdens. For example, in \cite{ISO2012_MPEGCDVS22806}, the authors proposed PQ-SIFT to quantize local descriptors with over a large vocabulary with 1 million centroids for 16 sub-segments.% Therefore, each descriptor needs more than 1 million comparison operations to find their nearest centers.

% During CDVS development, two kinds of representative compression algorithms based on vector quantization and transform are widely discussed and both of them achieve significant improvement on compression performance and computation efficiency compared with predecessors.  In early stage of CDVS, the Test Model under Consideration (TMuC) for Compact Descriptors for Visual Search (CDVS) \cite{ISO2012_MPEGCDVS12367} has proposed to employ tree structured vector quantization (TSVQ) \cite{ISO2012_MPEGCDVS24737} and product quantization \cite{jegou2011product} to make a compact descriptor for visual search. However, the heavy memory consumption, about 400 MBs, from a large visual vocabulary is a big and serious concern for visual search especially in the context of a mobile device, and the huge vocabulary also increase the quantization time via searching a particular code-vector from a large quantization table.

%\begin{figure}[t]
%\centering
%\includegraphics[width=8.0cm]{MSVQ.jpg}
%\caption{Flow chart of the two stage vector quantization (MSVQ). }
%\label{fig:MSVQ}
%\end{figure}

The Multi-Stage Vector Quantization (MSVQ) scheme \cite{ISO2012_MPEGCDVS24780} was adopted into the test model TM2.0 and significantly reduced the size of quantization tables. The MSVQ quantization consists of two stages, i.e., Tree Structured Vector Quantization (TSVQ) for the original raw descriptors at the 1st stage and the subsequent Product Quantization (PQ) for the residuals at the 2nd stage\iffalse, as illustrated in Fig.\ref{fig:MSVQ}\fi. After training the MSVQ over 6 million SIFT descriptors\iffalse extracted from the MIRFLICKR-25000 database \cite{huiskes08}\fi, the MSVQ only utilizes a 2-level tree-structure quantization table with 2048 visual centers and a PQ table with $\sim$16K centers. Therefore, in total comparison operations are reduced to 256 (1st level TSVQ)+8 (2nd level TSVQ)+16K (PQ) for each descriptor, which significantly reduces the computational cost in searching codewords.

% The Multi-Stage Vector Quantization (MSVQ) scheme \iffalse \cite{ISO2012_MPEGCDVS24780}\fi was adopted into the test model TM2.0 and significantly reduced the size of quantization table. The MSVQ quantization consists of two stages, i.e., Tree Structured Vector Quantization (TSVQ) for the original raw descriptors at the 1st stage and the subsequent Product Quantization (PQ) for the residuals at the 2nd stage\iffalse, as illustrated in Fig.\ref{fig:MSVQ}\fi. After training the MSVQ over 6 million SIFT descriptors extracted from the MIRFLICKR-25000 database \cite{huiskes08}, the MSVQ only consumes less than 5MBs memory, but it achieves the comparable matching and retrieval performance with TSVQ, which needs about 400 MBs. The smaller quantization table also reduces the quantization time in searching code-vectors.

%\begin{figure}[t]
%\centering
%\includegraphics[width=3.0cm]{hog.jpg}
%\caption{SIFT histogram of gradients $\bm{h}$. }
%\label{fig:HoG}
%\end{figure}

To further reduce the computational cost, CDVS finally adopts the transform coding with scalar quantization instead of the MSVQ. For each SIFT subregion Histogram of Gradients (HoG) $\bm{h}$ with bins \{$h_0\cdots h_7$\}\iffalse as shown in Fig.\ref{fig:HoG}\fi, CDVS applies an order-8 linear transform to capture the shape of HoGs. To improve the discriminative power of descriptors, CDVS defines two sets of transforms and applies different transforms to neighboring subregion HoGs, and the transforms are implemented via addtion/subtraction and shift operations with extremely low complexity. \iffalse as defined in Eqn.(\ref{Eqn_setA}) and Eqn.(\ref{Eqn_setB}),\fi
%\begin{equation}
%\left\{
%\begin{aligned}
%v_0 &= (h_2 - h_6)/2 \\
%v_0 &= (h_3 - h_7)/2 \\
%v_0 &= (h_0 - h_1)/2 \\
%v_0 &= (h_2 - h_3)/2 \\
%v_0 &= (h_4 - h_5)/2 \\
%v_0 &= (h_6 - h_7)/2 \\
%v_0 &= ((h_0 + h_4) - (h_2 + h_6))/4 \\
%v_0 &= ((h_0 + h_2 + h_4 + h_6) - (h_1 + h_3 + h_5 + h_7))/8
%\end{aligned}
%\right.
%,
%\label{Eqn_setA}
%\end{equation}
%\begin{equation}
%\left\{
%\begin{aligned}
%v_0 &= (h_0 - h_4)/2 \\
%v_0 &= (h_1 - h_5)/2 \\
%v_0 &= (h_7 - h_0)/2 \\
%v_0 &= (h_1 - h_2)/2 \\
%v_0 &= (h_3 - h_4)/2 \\
%v_0 &= (h_5 - h_6)/2 \\
%v_0 &= ((h_1 + h_5) - (h_3 + h_7))/4 \\
%v_0 &= ((h_0 + h_1 + h_2 + h_3) - (h_4 + h_5 + h_6 + h_7))/8
%\end{aligned}
%\right.
%.
%\label{Eqn_setB}
%\end{equation}
Each element of the transformed descriptors is further individually quantized to three values, $-1$, $0$ and $+1$, using quantization thresholds calculated from the off-line learned probability density function of that element. This transform coding method exhibits much lower computational cost than MSVQ while keeping comparable performance. \iffalse In addition, the scalar quantization only needs two comparison operations for each descriptor.\fi More importantly, the transform coding method is codebook free, and it is more suitable for GPU since the I/O speed is actually the bottleneck of GPU in large-scale computing.

\subsection{Local Feature Aggregation}\label{local_feature_aggregation}
The global descriptors with highly efficient distance computation are crucial for fast large-scale visual search. Different global descriptors were proposed in CDVS development such as Residual Enhanced Visual Vector (REVV)\cite{chen2011residual,ISO2012_MPEGCDVS23578}, Robust Visual Descriptor (RVD)\cite{husain2016improving,ISO2012_MPEGCDVS31426} and Scalable Compressed Fisher Vector (SCFV)\cite{lin2014rate,ISO2012_MPEGCDVS31401}. The REVV utilizes a set of 190 centroids from k-means clustering of SIFT descriptors off-line, and assigns each uncompressed SIFT descriptor of an input image to its nearest centroid in terms of L2 distance. The difference or residual between each SIFT descriptor and its nearest centroid is computed, and the mean residual of all the SIFT descriptors quantized to the same centroid is computed for a centroid. A power law with exponent value 0.6 is applied\iffalse to the values of the mean residuals\fi. With dimension reduction via PCA, these residual vectors are binarized according to its sign and concatenated to form a global descriptor. Although the REVV is with very low computational cost and memory footprint, the performance is yet to be improved.

\begin{figure*}[t]
  \centering
  \includegraphics[width=16cm]{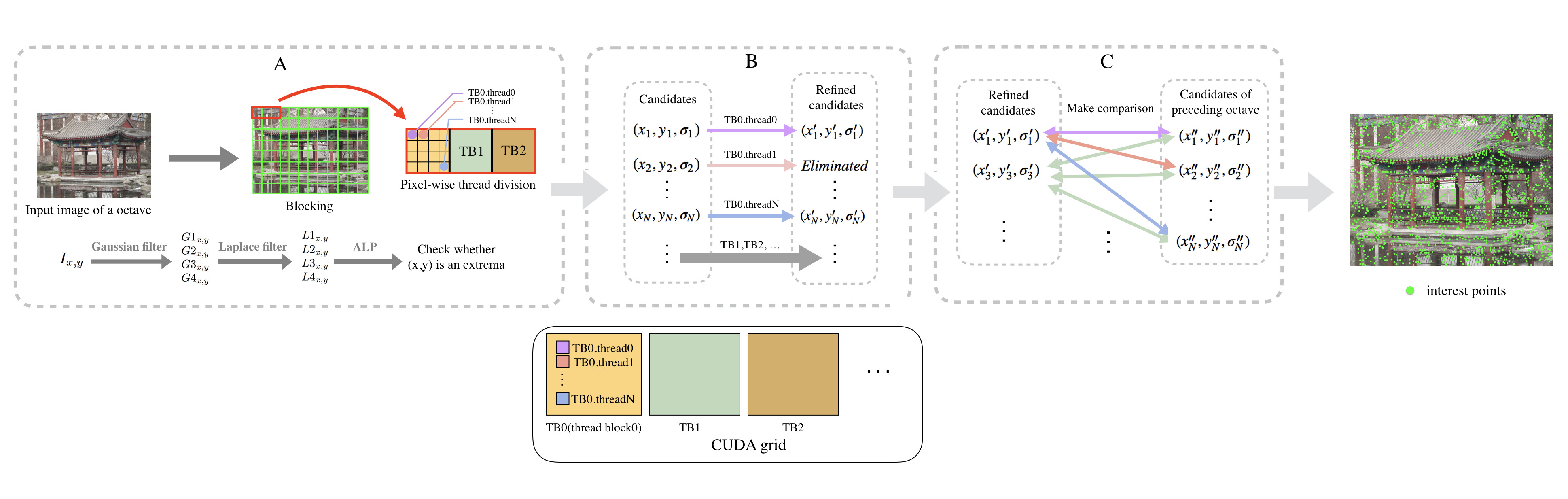}
  \caption{The block diagram of the interest point detection implementation on GPU}
\label{fig:Imp1}
\end{figure*}

Bober \emph{et al.} proposed an enhanced global descriptor, RVD, which improves the robustness of REVV by assigning each SIFT descriptor to multiple cetroids while reducing the computational cost. To reduce the computational cost, the RVD shifts the PCA to the first stage to reduce the computational cost for the subsequent calculations, and utilizes the matrix multiplication to implement PCA\iffalse and trains the PCA matrix from over 5 millions SIFT feature descriptors, and utilizes the matrix multiplication instead of the traditional PCA calculation\fi. This transform reduces the PCA computational complexity from $\mathcal{O}(n^2)$ to $\mathcal{O}(n)$, where $n$ is the number of local features. In addition, the matrix operations has been well implemented and optimized on GPU by NVIDIA, i.e., cuBLAS, and can greatly reduce the global memory operations compared with non-matrix operations. Furthermore, the RVD utilizes the L1 norm distance between SIFT descriptors and centroids instead of L2 norm distance in REVV to avoid the computation-intensive multiplication operations. Although each of the SIFT descriptors is assigned to multiple centroids with a bit increased computation, the RVD achieves better performance compared with REVV.

%. Before the aggregation, the residual vectors are first L1-Normalized, and then the weighted average of residual vectors in Eqn.(\ref{Eqn_RVD}) are cascaded and quantized as the global descriptors.
%\begin{equation}
%rv_j = \sum_{i=1}^{\emph{KN}}w_i \sum_{x_t \in \Omega_{k_j}} \frac{x_t-k_j}{|x_t-k_j|},
%\label{Eqn_RVD}
%\end{equation}
%where $x_t$ is the SIFT vector, $k_j$ is the $j^{th}$ centroid, $\Omega_{k_j}$ is the neighborhood of $k_j$ and $\emph{KN}=3$ and $w_1=4, w_2=2, w_3=1$. The RVD achieves better visual search performance with the limited increase of computational cost compared with REVV.

Finally, the CDVS adopted the SCFV as its global descriptor, which takes a Gaussian Mixture Model (GMM) with 512 components to capture the distribution of up to $N$ selected local feature descriptors. The SCFV also firstly reduces the SIFT dimensions utilizing PCA matrix, but it transforms the 128D SIFT into 32D instead of 48D of RVD, which further reduces the computational cost for subsequent operations.
%To be satisfy variant bandwidth requirements, SCFV selects a subset Gaussian components based on the standard deviation of 32-dimensional accumulated gradient vector $g = [g_0, g_1, . . . , g_{31}]$ as Eq.(\ref{Eqn:stdGMM}),
%This can be efficient implemented by multiplying a $32\times 128$ matrix for each local feature with low computational complexity $\mathcal{O}(F)$. To be satisfy variant bandwidth requirements, SCFV selects a subset Gaussian components based on the standard deviation of 32-dimensional accumulated gradient vector $g = [g_0, g_1, . . . , g_{31}]$ as Eq.(\ref{Eqn:stdGMM}),
For each 32D vector $x_t$, the major computation in SCFV include probability in Eqn.(\ref{Eqn:wprob}), the accumulated gradient vector $g_{\mu_i}^x$ with respect to the mean of the $i^{th}$ Gaussian function in Eqn.(\ref{Eqn:gradient}), and its standard deviation $\delta(i)$ in Eqn.(\ref{Eqn:stdGMM}),
\begin{equation}
\gamma_t(i) = \frac{w_ip_i(x_t|\lambda)}{\sum_{j=0}^{N}w_jp_j(x_t|\lambda)},
\label{Eqn:wprob}
\end{equation}

\begin{equation}
g_{\mu_i}^{x} = \frac{1}{K\sqrt{w_i}}\sum_{t=0}^{K-1}\gamma_t(i)\frac{x_t-\mu_i}{\sigma_i},
\label{Eqn:gradient}
\end{equation}

\begin{equation}
\delta(i) = \sqrt{\frac{1}{32}\sum_{j=0}^{31}(g_j-\frac{1}{32}\sum_{k=0}^{31}g_k)^2},
\label{Eqn:stdGMM}
\end{equation}
where $p_i(x_t|\lambda)$ is the Gaussian function, $w_i$ is the weight of the $i^{th}$ Gaussian function, and $N$ is the number of local features. Then, the Gaussian components are ranked in descending order according to $\delta$, and the top $K$ ones are selected based on the bit budget. Since only 250 local feature and 512 Gaussian functions are applied in CDVS, there are around 4.5 million multiplications/divisions for these 32D features, which is more than that in RVD, about 36 thousands multiplicaitons/divisions. However, the SCFV achieves very promising search performance and fulfills novel bit rate scalability, and moreover all these calculations can be transformed into matrix operations, which can be well implemented and optimized in parallel on GPU achieving significant speedup. The detailed speedup results for SCFV on GPU are shown in Fig.\ref{fig:modules_gpu_c}.

\section{Fast CDVS Encoder using GPU-CPU Hybrid Computing}

Although great efforts have been made to reduce the computational complexity of CDVS, it is still difficult to implement highly efficient CDVS on CPU even with multiple threads in parallel. By leveraging the massive parallel process cores of GPU, we design and implement the very fast CDVS encoder using GPU-CPU hybrid computing. Three major time-consuming modules, i.e., interest point detection, local feature description and aggregation, are shifted to GPU platform, while the others remains on CPU platform as shown in Fig.\ref{fig:NormBlockCDVS}. In addition, the local feature compression and aggregation are independent process, they can be in parallel performed on CPU and GPU simultaneously, which elegantly leverages the computational resources on CPU and GPU. In the following subsection, we present technical details on interest point detection, local feature description and aggregation.

% Although the local feature compression consumes more computational time compared with local feature aggregation, it is a highly data dependent process, especially for the arithmetic coding process, which is not suitable for GPU implementation.

\subsection{Interest Point Detection on GPU}

In interest point detection, we adopt sperate Gaussian filters to construct the image pyramid, utilize Laplacian filter and ALP to detect interest points. Distinct from BFLoG, the spatial domain Gaussian and Laplacian filtering can be implemented in much higher degree of parallelism with very low memory usage, while the BFLoG is only implemented on block-level parallel with doubling memory usage due to the complex operations in FFT. In addition, the spatial domain filtering can be completed in one step, while the BFLoG need three sequential steps, i.e., FFT, element-wise production and IFFT, which incurs more process time for GPU.

The implementation of interest point detection comprises of three basic stages as illustrated in Fig.\ref{fig:Imp1}. Part A illustrates the implementation of LoG filtering and ALP detection, which outputs the interest point candidates. To achieve high degree parallelism, the input image is first divided into $N\times N$ blocks, and each block is assigned to a thread block (TB) to perform LoG filtering. Since the access speed for shared memory is much higher than graphic memory, these image blocks are loaded into the shared memory of their corresponding thread blocks. Then, the threads in the same TB access the pixels from its shared memory sequentially, i.e., TB0.thread0 access the first pixel, TB.thread1 access the second pixel, and so no. This memory access mechanism is to make full use of the merit of \textit{global memory coalescing} to reduce memory access time cost. Then, the Gaussian filtering, Laplacian filtering and ALP detector are sequentially performed in corresponding threads. In our implementation, we specify that both image block and thread block are of the same size, in which each pixel is processed on each thread in parallel.

\begin{figure*}[t]
  \centering
  \includegraphics[width=14cm]{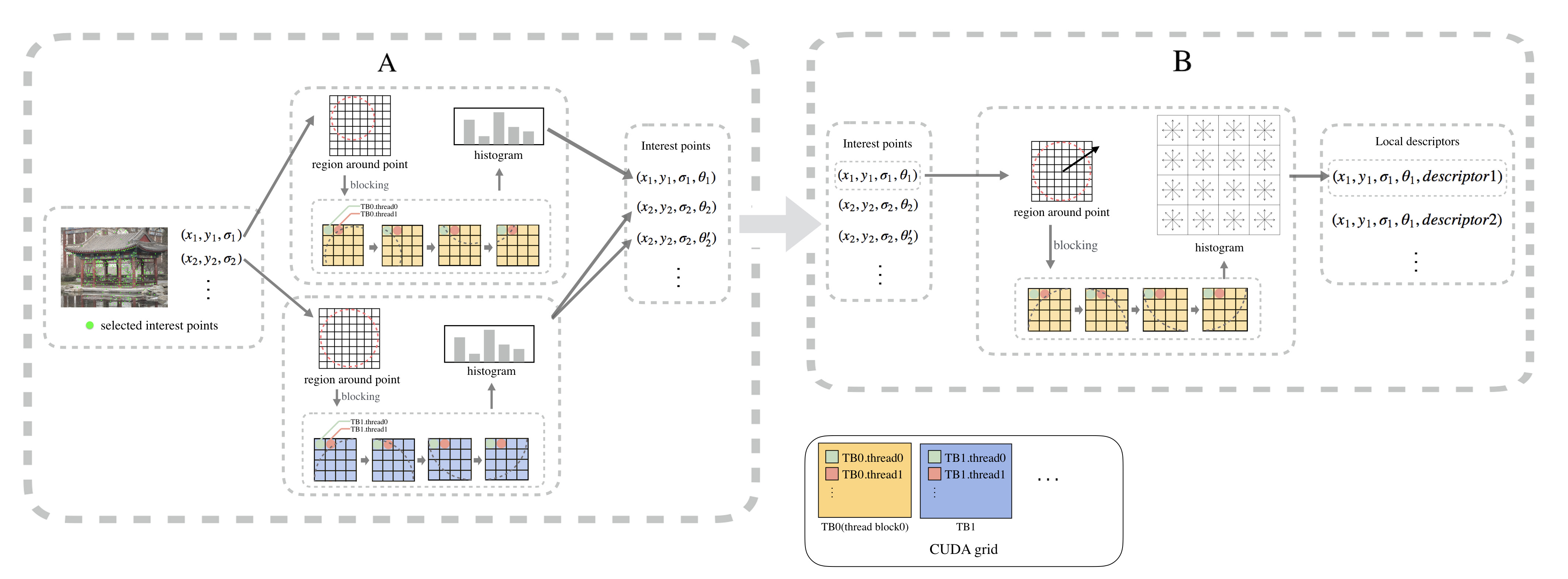}
  \caption{The block diagram of the local feature description implementation on GPU}
\label{fig:Imp2}
\end{figure*}

For Part B, the interest point candidates are refined to remove unstable interest points and more accurate locations of interest points are determined. To speed up the process, we reorganize the interest point data $(x_i,y_i,\sigma_i)$ and the corresponding pixels in its $3\times 3$ neighborhood into a continuous queue. For $N$ continuous candidates, we assign a TB with $N$ threads to calculate the LoG response in a $3\times 3$ region around the candidate, which employs the global memory coalescing to speed up the process. Afterwards, the interest points considered as an unstable ones are removed out, otherwise, more accurate position are updated by the interpolation.

In Part C, the final interest points are determined by comparing the neighboring ones in adjacent octaves. Likewise, we first construct two queues for the interest point candidates in current octave and preceding octave respectively, and then apply one TB to perform the comparisons between one candidate in current octave and all the candidates in preceding octave by the global memory coalescing. Each thread first executes the comparison between two candidates. If they are close enough in $(x,y,\sigma)$ space, the candidate with lower LoG response is eliminated. The implementation details on each thread are illustrated in \textbf{Algorithm \ref{InterestPointDetectionAlgo}} using  pseudo-code.

\renewcommand{\algorithmicrequire}{\textbf{Input:}}
\renewcommand{\algorithmicensure}{\textbf{Output:}}
\begin{algorithm}[htp]
      \caption{Interest Point Detection on GPU}
      \label{InterestPointDetectionAlgo}
      \begin{algorithmic}[1]
        \Require
          Input image $I$, $G_{0,0}=I$
        \Ensure
          A queue of interest points
        \For{$o$ = 0 to $num\_octave$}
        		\State //Stage A: Scale-space for $G_{o,0}$
        		\State Load $G_{o,0}$ to shared memory
        		\ParFor{each pixel($i$,$j$)}
        			\State //A thread corresponds to a pixel
            		\State $T_{k}[i][j]$ = $\sum_{p=-r_k}^{r_k}G_{o,0}[i+p][j]*g_k[p]$
            		\State $G_{o,k}[i][j]$ = $\sum_{p=-r_k}^{r_k}T_k[i][j+p]*g_k[p]$
            		\State $L_{o,k}[i][j] = \sum_{m,n=-1}^{1}G_{o,k}[i+m][j+n]*f[m][n]$
                    \State Local extremum determination via Eqn.(2)
%            		\State Local extremum scale $\sigma$ determination via Eqn.(2)
%            		\State Calculate LoG response $p(i,j,\sigma)$ using Eqn.(2)
%            		\If {$p(i,j,\sigma)$ is the extrema in the $3\times 3$ region}
%                		\State put $p(i,j,\sigma)$ into $Q_{o}$ with atomic operation
%            		\EndIf
        		\EndParFor

            \State //Stage B: Refine extremum location
            	\ParFor{each candidate $(x_o,y_o,\sigma_o)$ in $Q_{o}$}
            		\State //A thread corresponds to a candidate
            		\State $(x'_o,y'_o,\sigma'_o)$ = CoordinateRefine$(x_o,y_o,\sigma_o)$
%            		\If {$(x'_o,y'_o,\sigma'_o)$ is at local edges or unstable}
%                		\State Eliminate the $(x_o,y_o,\sigma_o)$ from $Q_o$
%            		\Else
%                		\State Replace $(x_o,y_o,\sigma_o)$ with $(x'_o,y'_o,\sigma'_o)$ in $Q_o$
%            		\EndIf
            \EndParFor

            \State //Stage C: Extremum determination between octaves
            %\ParFor{each two interest points from different octaves: $(x'_o,y'_o,\sigma_o)$ and $(x'_{o-1},y'_{o-1},\sigma'_{o-1})$}
            \ParFor{ a pair candidates }
            		\State //A thread corresponds to a pair
            		\State $dis$ = distance($(x'_o,y'_o,\sigma_o)$,$(x'_{o-1},y'_{o-1},\sigma'_{o-1})$)
                    \State Update $Q_o$ and $Q_{o-1}$ according $dis$ and $p(x'_o,y'_o,\sigma'_o)$, $p(x'_{o-1},y'_{o-1},\sigma'_{o-1})$
%            		\If {$dis < threshold$  }
%                		\If {$p(x'_o,y'_o,\sigma'_o) < p(x'_{o-1},y'_{o-1},\sigma'_{o-1})$}
%                			\State Eliminate $(x'_o,y'_o,\sigma'_o)$ from $Q_o$
%                		\Else
%                    		\State Eliminate $(x'_{o-1},y'_{o-1},\sigma'_{o-1})$ from $Q_{o-1}$
%            			\EndIf
%            		\EndIf
            \EndParFor
			\State Merge $Q_o$ into output
			\State Downsampling $G_{o,4}$ to $G_{o+1,0}$
        \EndFor

     \end{algorithmic}
\end{algorithm}

The pixel-level thread allocation and efficient shared memory assess significantly reduce runtime cost for interest point detection. \iffalse Since there is no data dependency in these operations, our implementation on GPU can get the identical calculation results as that in CPU except for machine precision.\fi Based on our experimental results, the optimal GPU implementation can achieve around 26 times speedup compared with that on CPU platform, and the running time decrease from 56.2ms to 2.16ms for $640\times 480$ images.

\subsection{Local Feature Description on GPU}
The local feature description includes two stages, i.e., orientation computation and SIFT description. To allow rotation invariance for local feature descriptors, each interest point is assigned one or more dominant orientations based on the distribution of the quantized gradient directions in its neighborhood with radius $3.96\sigma$. To derive the dominant orientation, an orientation histogram with 36 bins shall be formed from the computed gradient orientations. The orientations with bin values greater than 0.8 times of the highest peak are kept as orientations of the interest point. For SIFT description, the image patch centered at interest point $(x,y)$ is first rotated to the angle of its orientation, and then it is divided into 4 horizontal and 4 vertical spatial subregions referred to as cells. The size of each side of each cell shall be $3\sigma$ pixels. From each cell, a histogram of gradients with 8 orientation bins, referred to as cell histogram, is generated. The SIFT descriptor is formed by concatenating these cell histograms.

Based on the above introduction, the local feature description is divided into two stages as illustrated in Fig.\ref{fig:Imp2}. The histogram construction occupies major computation. To maximize the degree of parallelism, we assign one TB to each interest point to generate the orientation histogram and compute the dominant orientations. However, the pixel-level parallelism with global memory coalescing is difficult to implement by simply allocating one thread to compute the gradient of each pixel and sum them up into a histogram. This is because that the number of threads in one TB should be the same, while the image patch size $3.96\sigma$ is variant for different interest points. If we allocate threads according to the largest image patch, the amount of pixels is much more than the available threads in one TB.

\begin{algorithm}[htp]
      \caption{Local Feature Description on GPU}
      \label{LocalFeatureDescriptionAlgo}
      \begin{algorithmic}[1]
        \Require
          Interest points after feature seletction
        \Ensure
          The descriptors of interest points
        %\State //The max size of a region is $R*R$
        \State //Stage A: orientation assignment
        \ParFor{each interest point $(x_{ctr},y_{ctr},\sigma)$ }
		\State //Apply a thread block to an interest point
		%\State $r_{ori} =3.96*\sigma$
		\State Load $R\times R$ regions in shared memory
		%\State //Divide the $R*R$ region into $N*N$ blocks
		%\ParFor{each $(tx,ty)$ in $[0..N,0..N]$}
        \ParFor{each $(x,y)$ in $R\times R$}
			\State //a thread corresponds to a pixel
			    \If {distance$((x_{ctr},y_{ctr})$,$(x,y)) < 3.96\sigma$  }
%						\State $l$ = GradientLength($x,y$)
%					    \State $a$ = GradientAngle($x,y$)
%					    \State Add ($l,a$) to histogram with atomicadd
                    \State Add gradient magnitudes and orientations to histogram
			    \EndIf
			%\State Synchronize()
%			\If {$(tx,ty) == (0,0)$  }
%			    \State // The first thread in a thread block
				\State Compute the orientation from the histogram
%			\EndIf
		\EndParFor
		\EndParFor

        \State //Stage B: Feature description

        \ParFor{Each interest point $(x_{ctr},y_{ctr},\sigma,\theta)$ }
		\State //Apply a thread block to an interest point
		%\State $r_{ori} =6*\sigma$
		%\State //Divide the $R*R$ region into $N*N$ blocks
		%\ParFor{each $(tx,ty)$ in $[0..N,0..N]$}
		%	\State //a thread corresponds to a (tx,ty)
		%	\For{each $(i,j)$ in $[0..R/N-1,0..R/N-1]$}
		%	    \State $x = i*N + tx + int(x_{ctr} - r_{ori})$
		%	    \State $y = i*N + ty + int(y_{ctr} - r_{ori})$
            \ParFor{each $(x,y)$ in $R\times R$}
            \State //a thread corresponds to a pixel
			    \If {distance$((x_{ctr},y_{ctr})$,$(x,y)) < 6\sigma$  }
				%		\State $l$ = GradientLength($x,y$)
%					    \State $a$ = GradientAngle($x,y$)-$\theta$
					   % \State Add ($l,a$) to histogram with atomicadd
                     \State         Add gradient magnitudes and orientations to histogram
			    \EndIf
			\EndParFor
			%\State Synchronize()
%			\If {$(tx,ty) == (0,0)$  }
%			    \State // The first thread in a thread block
                \State Generate the 128D descriptor from the histogram
%				\State Generate the 128-dim descriptor from the histogram and normalize it
%			\EndIf
		%\EndParFor
		\EndParFor
     \end{algorithmic}
\end{algorithm}

To solve the problem of variant patch size, we design a novel block-based pixel-level parallelization method. For each image patch, we divide it into non-overlapped $N\times N$ sub-patches, and allocate each TB with $N\times N$ threads. Then, the gradient computation can be performed in pixel-level parallelism, and gradients are then exported to shared memory to form gradient histograms by atomic addition. The detailed design for local feature description is illustrated via pseudocode in \textbf{Algorithm \ref{LocalFeatureDescriptionAlgo}}. Compared with the feature-level parallelization which assigns one independent thread to each interest point, the proposed block-based pixel-level parallelization is more efficient. Because feature-wise parallelization would lead to very unbalanced workload among threads and thus degenerate the processing efficiency on GPU due to variant patch sizes for interest points. Besides, local feature selection is applied in CDVS to keep less than $N$ ($N$ is usually smaller than 650) features for describing an image. It means that at most $N$ threads can be launched simultaneously for feature-wise parallelization, which incurs much less than cores in GPU. Our GPU implementation for local feature description achieves more than 90 times speedup compared that on CPU at different bitrates, the running time significantly decreases from $33.2\sim 79.2$ ms to $0.4\sim 0.83$ ms.

\subsection{Local Feature Aggregation on GPU}

When the stages of interest point detection and local feature description have been implemented optimally, the proportion of time-consumption for local feature aggregation increase from 10\% up to 80\%, which becomes the bottleneck in the whole pipeline of CDVS encoding. Hence, the remaining issue is to speed up the local feature aggregation. As introduced in Section \ref{local_feature_aggregation}, there are two parts, i) dimension reduction via PCA, ii) the fisher vector aggregation. The PCA calculation in CDVS is defined by matrix multiplication, which can be well fulfilled by calling the highly optimized matrix operation lib in CUDA.

To speed up the fisher vector aggregation, we first transform the probability calculation for each 32D local descriptors in Eq.(\ref{Eqn:prob}) into the matrix multiplications and a set of element-wise operations as shown in Egn.(\ref{Eqn:probM}).
\begin{equation}
P_{i,j} = \sum_{k=0}^{31} \frac{(D_{i,k}-M_{j,k})^2}{V_{j,k}},
\label{Eqn:prob}
\end{equation}

\begin{equation}
P = (D.*D)*(1./V^T)-2D*(M./V)^T + O*((M.*M)./V)^T.
\label{Eqn:probM}
\end{equation}
Here, $P_{i,j}$ represents the probability of the $i^{th}$ descriptor under the $j^{th}$ GMM component, $D_{i,k}$ denotes the $k^{th}$ dimension of the $i^{th}$ descriptor, $M_{j,k}$ and $V_{j,k}$ denote the $k^{th}$ dimension of mean and covariance vector for the $j^{th}$ GMM component, O is a matrix filled by 1 and has the same dimension with $D$. The notations, '.*' and './' represent the element-wise multiplication and division. These matrix operations can be efficiently performed by calling the well optimized matrix operation library in CUDA. Similarly, we derive the matrix implementation for the calculation of accumulated gradient vectors with respect to the mean and variance denoted as $(GM)$ and $(GV)$, the calculations of which are transformed from Eqn.(\ref{Eqn:gvm}) and Eqn.(\ref{Eqn:gvv}) to Eqn.(\ref{Eqn:gvmM}) and Eqn.(\ref{Eqn:gvvM}).
\begin{equation}
GM_{i,j} = \sum_{k=0}^{DescNum}(\frac{D_{i,k}- M_{j,k}}{V_{j,k}} * Q_{i,j}),
\label{Eqn:gvm}
\end{equation}

\begin{equation}
GM = (Q^T*D - Q^T * O .* M)./V,
\label{Eqn:gvmM}
\end{equation}

\begin{equation}
GV_{i,j} = \sum_{i,j}^{DescNum}((\frac{D_{i,k}-M{j,k}}{V_{j,k}})^2-1)*Q_{i,j},
\label{Eqn:gvv}
\end{equation}
\begin{equation}
\begin{aligned}
GV &= (Q^T*(D.*D) - Q^T*D.*2M \\
& + Q^T*O.*(M.*M-V.*V)) ./ (V.*V).\\
\end{aligned}
\label{Eqn:gvvM}
\end{equation}
Here $Q$ is the normalized probability of $P$, and $DescNum$ is the number of the local feature descriptors. After these conversions, local feature aggregation module can be implemented by invoking the matrix lib in CUDA, which is well optimized by NVIDIA. Based on the experimental results, the running time of local feature aggregation decreases from around 13 ms to 0.25 ms, and the speedup of GPU implementation is more than 49 times compared that on CPU.

\section{Experimental Results and Analysis}

\subsection{Databases and Evaluation Criteria}

To analyze the performance of the fast CDVS encoder, we perform pairwise matching and image retrieval tasks on patch-level and image-level databases. Two image-level databases are utilized in our experiments,  1) MPEG-CDVS benchmark database \cite{ISO2011_MPEGCDVS}, which consists of 5 classes: graphics, paintings, video frames, landmarks and common objects; 2) \textit{Holiday} database \cite{jegou2008hamming}, which contains images from different scene types to test the robustness to transformations such as rotation, viewpoint, illumination changes and blurring, and is widely used in academic literatures. In addition, we also utilize two patch-level databases, MPEG-CDVS patch database \cite{Vijay2013_PatchCDVS} and Winder and Brown database \cite{winder2009picking}, which contain 100K and 500K matching pairs of $64\times 64$ pixels image patches involving canonical scaled and oriented patches.

For performance validation, we adopt the Mean Average Precision (mAP), True Positive Rate (TPR) and False Positive Rate (FPR) to measure the image retrieval and pair matching performance respectively. The mAP for a set of queries is calculated as the mean of the average precision scores for each query, which is defined as follows,
\begin{equation}
mAP = \frac{\sum_{q=1}^{Q}AP(q)}{Q},~AP = \int_{0}^{1} \bm{p}(r)dr,
\label{Eqn:mAP}
\end{equation}
where $Q$ is the number of queries, and $AP$ is the average precision,
%\begin{equation}
%AP = \int_{0}^{1} \bm{p}(r)dr,
%\label{Eqn:AP}
%\end{equation}
and $\bm{p}(r)$ is the precision function at recall $r$. The TPR and FPR are calculated as,
%\begin{equation}
%\emph{TPR} = \frac{\emph{TP}}{\emph{TP}+\emph{FN}},
%\label{Eqn:TPR}
%\end{equation}
%
%\begin{equation}
%\emph{FPR} = \frac{\emph{FP}}{\emph{FP}+\emph{TN}},
%\label{Eqn:FPR}
%\end{equation}
\begin{equation}
\emph{TPR} = \frac{\emph{TP}}{\emph{TP}+\emph{FN}},~\emph{FPR} = \frac{\emph{FP}}{\emph{FP}+\emph{TN}},
\label{Eqn:TPR_FPR}
\end{equation}
where \emph{TP}, \emph{FP}, \emph{TN} and \emph{FN} are the number of the true positive, false positive, true negative and false negative retrieval results.

The fast CDVS encoder is implemented based on the latest CDVS reference software TM14.0 using GPU-CPU hybrid computing. In the following section, we validate the improvements of visual search accuracy and descriptor extraction speed, respectively. \iffalse Since there are some non-normative fast algorithms for CDVS, we also compare our CDVS encoder (denoted CDVS\_GPU) with the reference software and the optimized reference software using non-normative fast algorithms on CPU platform, which are denoted as CDVS\_CPU and OPT\_CDVS\_CPU, respectively.\fi To fully understand the superior performance of CDVS, we also compare CDVS descriptor with state-of-the-art visual feature descriptors including SIFT \cite{lowe2004distinctive}, SURF \cite{bay2008speeded}, ORB \cite{Rublee_ORB2011}, BRISK \cite{Leutenegger_2011}, AKAZE \cite{Alcantarilla2013Fast} and LATCH \cite{Levi_2016}, which are from OpenCV 3.2.0 with default parameters.

%\begin{figure}[t]
%\centering
%\includegraphics[width=8.0cm]{mpegdataset1.jpg}
%\caption{Example images from the MPEG-7 CDVS datasets (from top to bottom: graphics, paintings, video frames, buildings and objects).}
%\label{fig:dataset}
%\end{figure}

\subsection{Performance Comparison}

\begin{figure}[t]

  \begin{minipage}[b]{0.235\textwidth}
  \centering
\subfigure[\emph{MPEG-CDVS patch database}]{\label{fig:patchroc_a}\includegraphics[width=4.2cm]{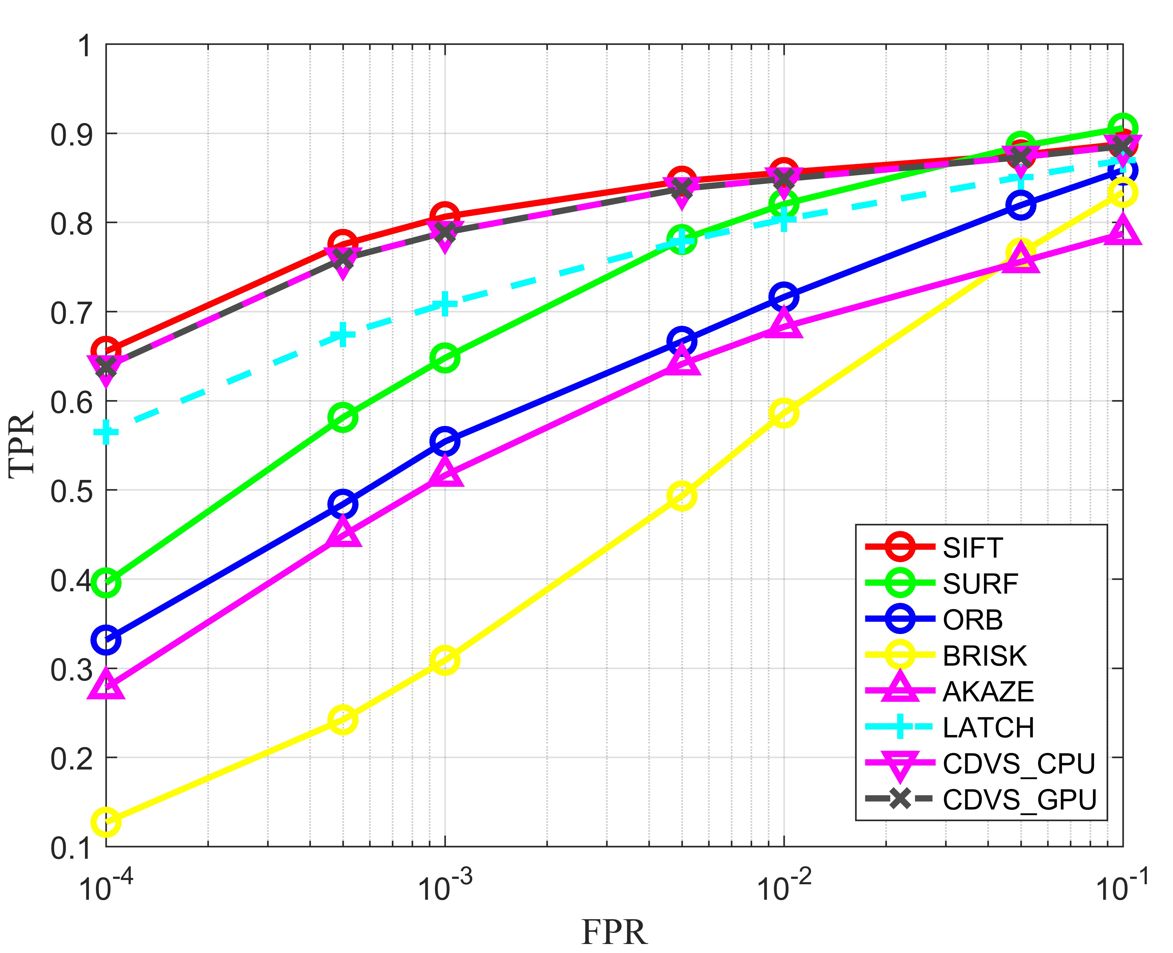} }
  \end{minipage}
  \hspace{0.5cm}
  \begin{minipage}[b]{0.235\textwidth}
  \centering
  \subfigure[\emph{Winder and Brown patch database}]{\label{fig:patchroc_b}\includegraphics[width=4.2cm]{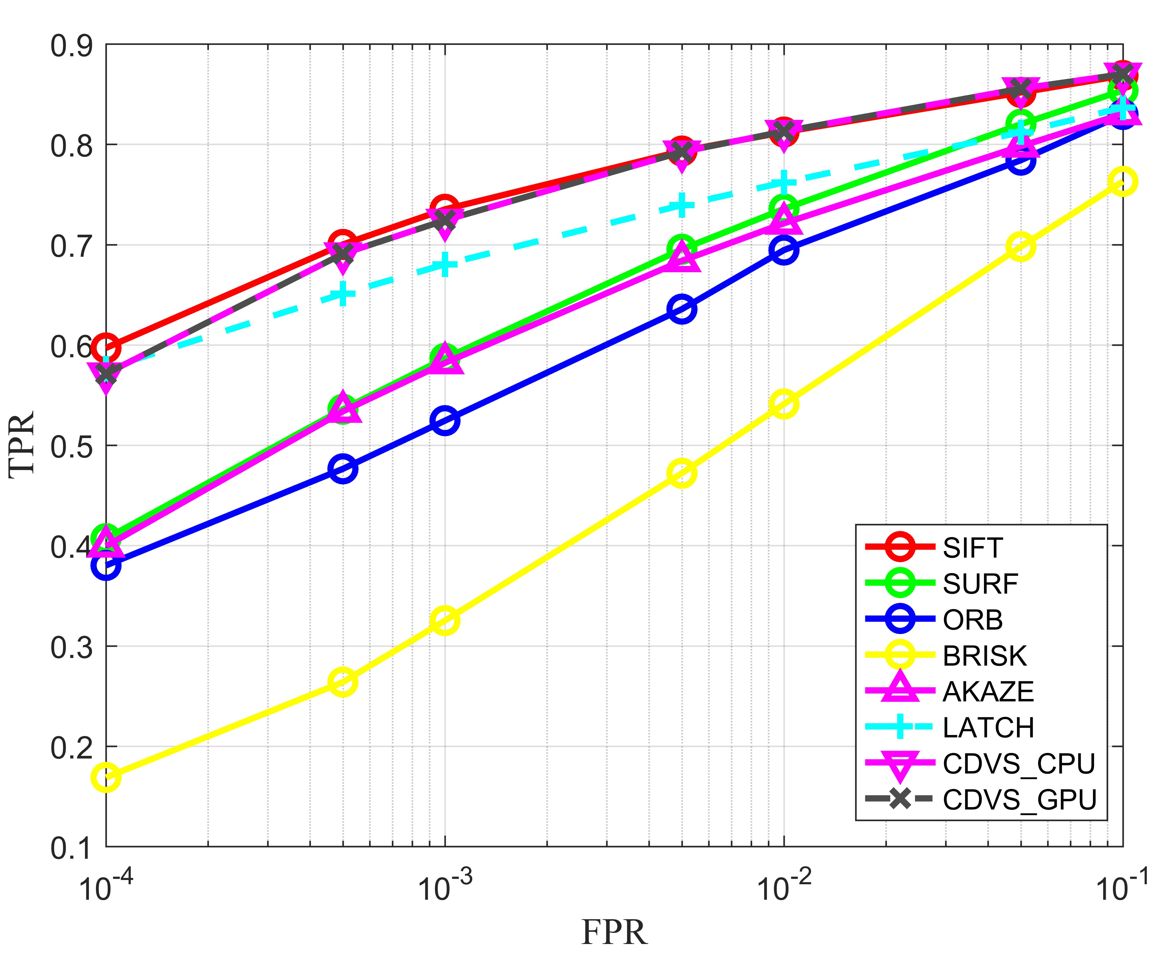} }
  \end{minipage}
  \caption{Comparison of ROC curves at patch level for different visual descriptors on (a) MPEG-CDVS database, (b) Winder and Brown database.}
\label{fig:patchlevel_roc}
\end{figure}

%To evaluate descriptors, we compare the CDVS descriptors generated on GPU and CPU platforms with the widely used visual search descriptors including SIFT \cite{lowe2004distinctive}, SURF \cite{bay2008speeded}, ORB \cite{Rublee_ORB2011}, BRISK \cite{Leutenegger_2011}, AKAZE \cite{Alcantarilla2013Fast} and LATCH \cite{Levi_2016}, which are all invoked from OpenCV 3.2.0 with default parameters. For CDVS descriptors, a Gaussian smoothing filter with $\sigma = 2.7$ is applied to the patches as suggested in \cite{Chandrasekhar2012}.

In this section, we first show the patch-level evaluation as it provides the initial idea about the performance of descriptors and avoids the influence of different interest point selection strategy. To understand the matching accuracy at different FPR, the ROC curves on MPEG-CDVS and Winder and Brown patch databases are illustrated in Fig.\ref{fig:patchlevel_roc}. The CDVS descriptors generated from CPU and GPU platforms (denoted as CDVS\_CPU and CDVS\_GPU) yields almost the same pairwise matching results, which verifies that the CDVS encoder is exactly implemented on heterogeneous architectures without performance sacrifice. The patch matching performance of CDVS is only inferior to the original SIFT descriptors, and much more superior to other visual feature descriptors. It is worthy to note, CDVS only needs $32\sim205$ bits for each descriptor under different rate configurations, and its bitrate is much fewer than that of SIFT, which costs 512 bytes for each descriptor. Although SURF, ORB, BRISK, AKAZE and LATCH are also compact descriptors, their performances are obviously inferior to CDVS and also cannot outperform CDVS in compactness. Herein, there are 256, 128, 256, 244 and 128 bytes in representing each SURF, ORB, BRISK, AKAZE and LATCH descriptor, respectively. In Fig.\ref{fig:patchlevel_roc}, we only utilize 103 bits for each CDVS descriptor.

\begin{figure}[t]

  \begin{minipage}[b]{0.235\textwidth}
  \centering
\subfigure[\emph{MPEG-CDVS database}]{\label{fig:patchroc_a}\includegraphics[width=4.2cm]{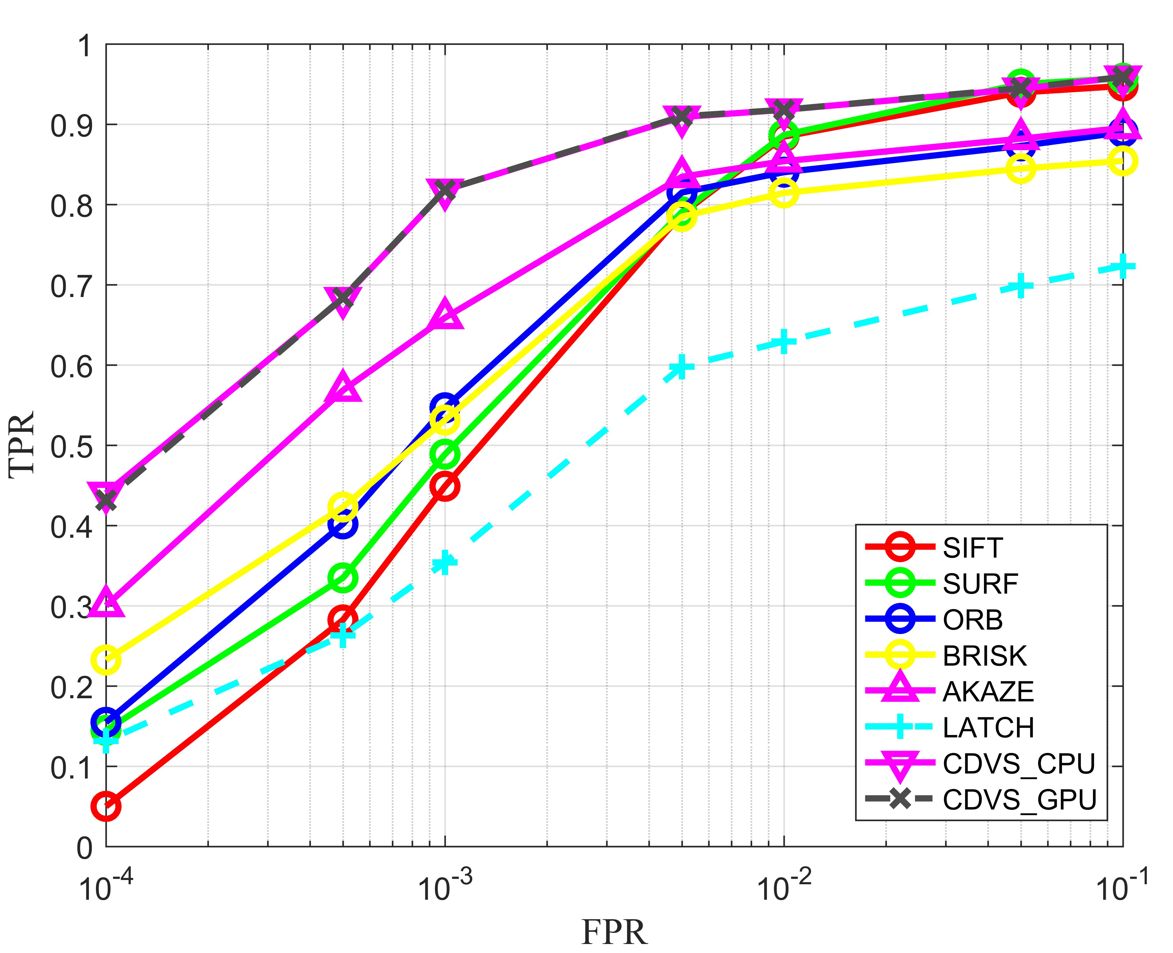} }
  \end{minipage}
  \hspace{0.5cm}
  \begin{minipage}[b]{0.235\textwidth}
  \centering
  \subfigure[\emph{Holiday database}]{\label{fig:patchroc_b}\includegraphics[width=4.2cm]{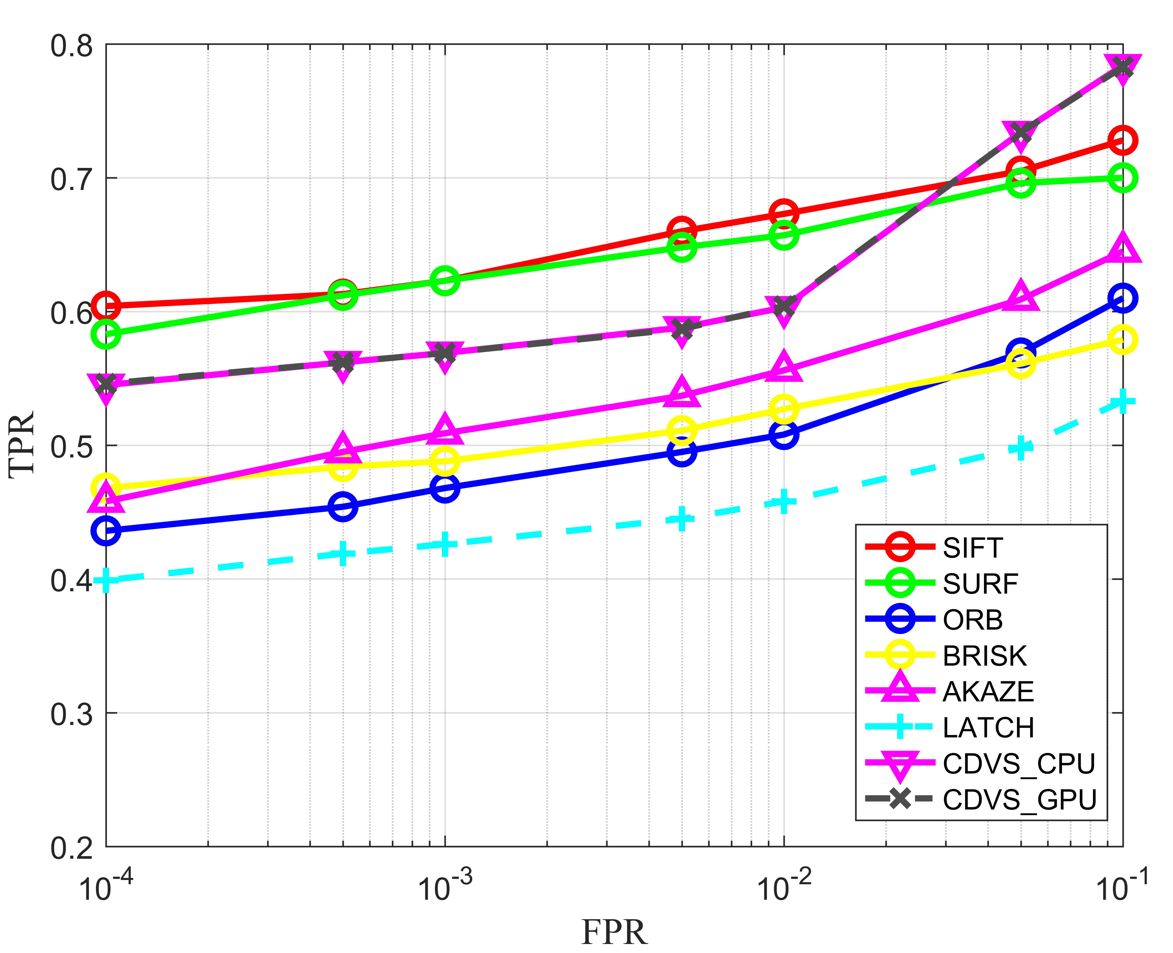} }
  \end{minipage}
  \caption{Comparison of ROC curves of image retrieval for different visual feature descriptors on (a) MPEG-CDVS database, (b) Holiday database.}
\label{fig:databaselevel_ROC}
\end{figure}

\begin{figure}[t]
\centering
%\begin{minipage}[b]{0.435\textwidth}
\includegraphics[width=8.0cm]{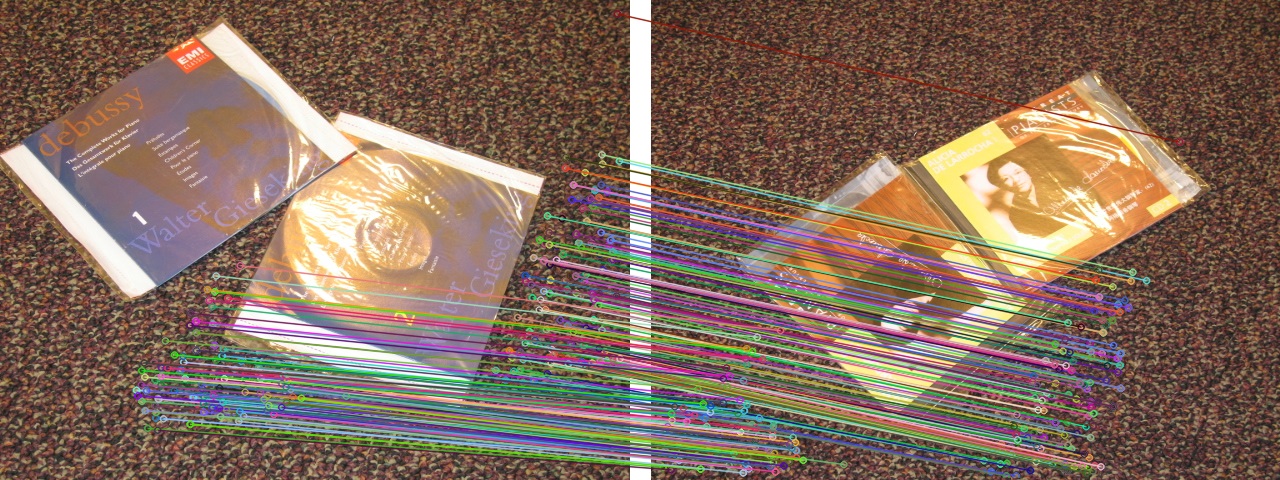}
%\subfigure[]{\label{fig:ratemodel_b}\includegraphics[width=7.0cm]{DistortionReduction.jpg} }
%\end{minipage}
\caption{The mismatching example using SIFT descriptors. }
\label{fig:falsematchExample}
\end{figure}

To verify the overall performance on real images, we carry out image-level evaluation for pair matching and image retrieval tasks. In pairwise matching database, there are 10,155 matching image pairs and 112,175 non-matching in MPEG-CDVS database, and 2072 matching image pairs and 20874 non-matching in Holiday database. Fig.~\ref{fig:databaselevel_ROC} shows the ROC curves on image pair matching task for different visual descriptors. The CDVS descriptors extracted using CPU and GPU platform achieve almost the same performance, and obviously outperform the other competitors. Remarkably, the performance of SIFT descriptors is poor when FPR is low. It is observed that, there are many false matched image pairs where the resulting matched SIFT points are on the background as illustrated in Fig.\ref{fig:falsematchExample}. This phenomena verifies that the local feature selection not only reduces the computational cost, but removes the non-meaningful interest points, which significantly contributes to high performance.

\begin{figure}[t]

  \begin{minipage}[b]{0.235\textwidth}
  \centering
\subfigure[\emph{MPEG-CDVS database}]{\label{fig:patchroc_a}\includegraphics[width=4.2cm]{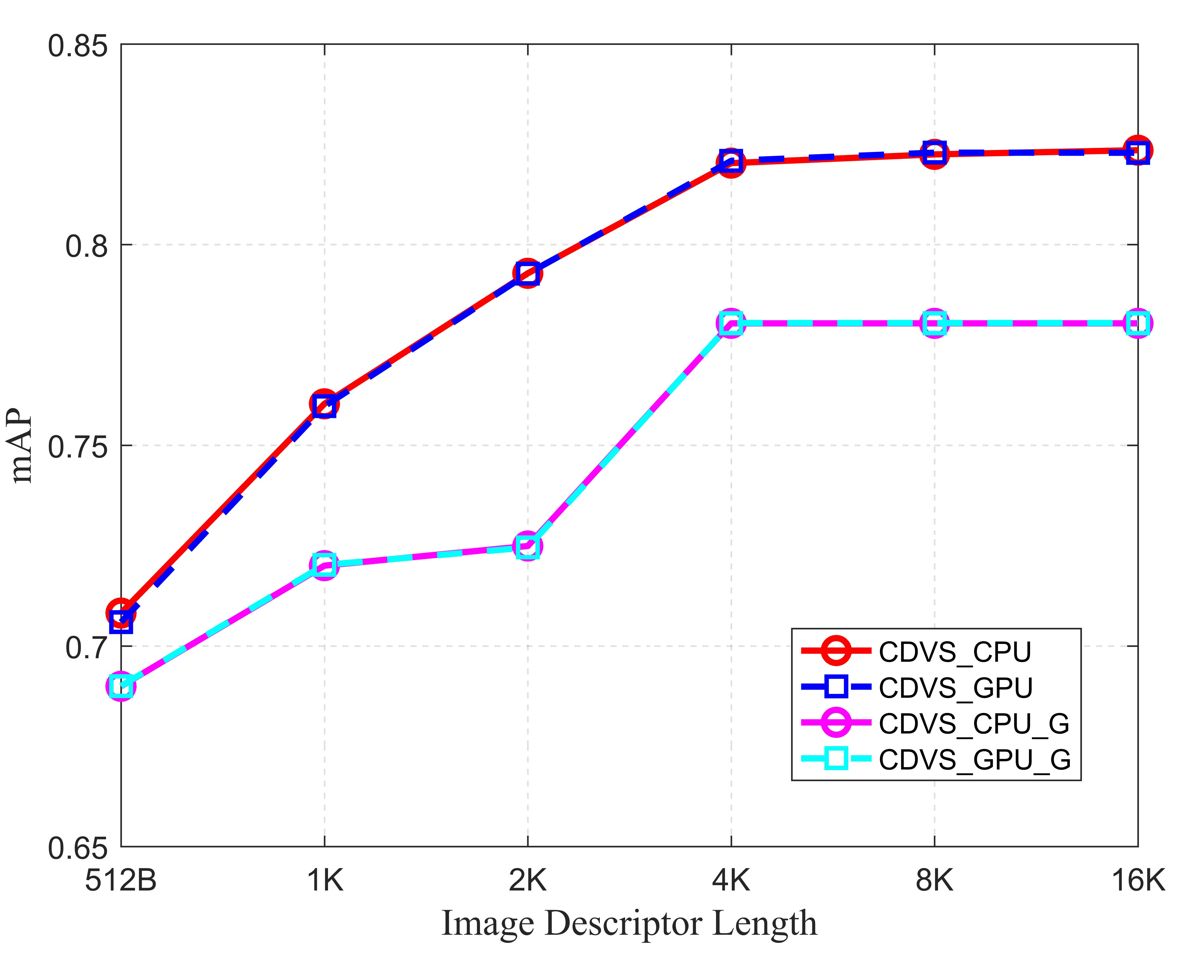} }
  \end{minipage}
  \hspace{0.5cm}
  \begin{minipage}[b]{0.235\textwidth}
  \centering
  \subfigure[\emph{Holiday database}]{\label{fig:patchroc_b}\includegraphics[width=4.2cm]{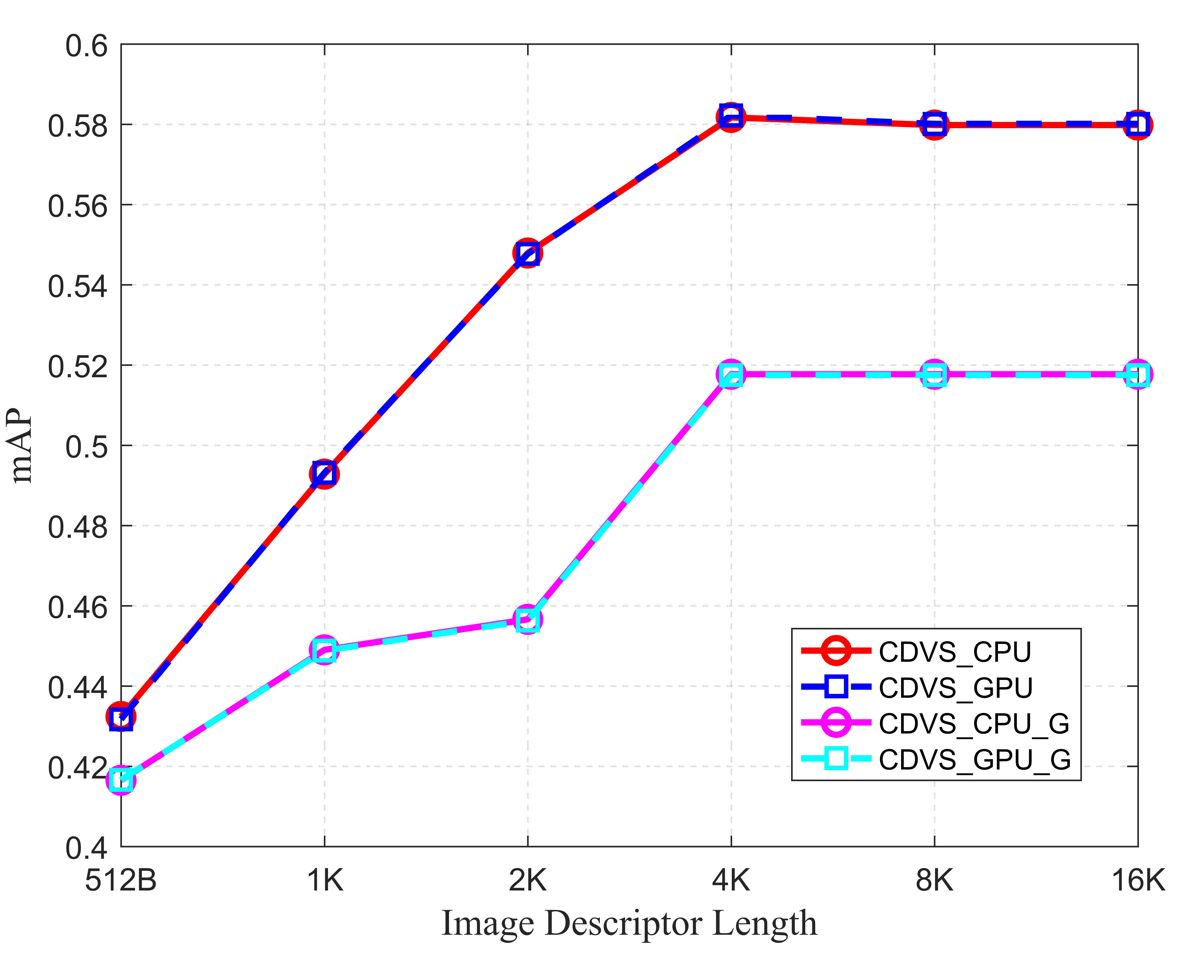} }
  \end{minipage}
  \caption{The mAP comparison of image retrieval using CDVS on CPU and GPU platform on (a) MPEG-CDVS database, (b) Holiday database. Herein, CDVS\_CPU\_G and CDVS\_GPU\_G represent the CDVS global descriptors generated on CPU and GPU platforms, repspectively.}
\label{fig:databaselevel_mAP}
\end{figure}

In the image retrieval experiments, we compare the performance of the CDVS and its global descriptors generated at 6 pre-defined
descriptor lengths: 512 bytes, 1K, 2K, 4K, 8K and 16K in MPEG common test conditions. From the results in Fig.\ref{fig:databaselevel_mAP}, we can see that the fast CDVS encoder is well implemented in parallel using GPU platform with negligible performance difference. \iffalse However, the OPT\_CDVS\_CPU with fast algorithms, e.g., BFLoG, introduce cause performance loss to some extent. \fi Table \ref{tab:mAP_CDVS} shows the detailed numerical mAP results for CDVS\_CPU \iffalse, OPT\_CDVS\_CPU \fi and CDVS\_GPU respectively, from which the same conclusion can be drawn.\iffalse In addition, the OPT\_CDVS\_CPU brings about xx\%$\sim$ xx\% performance loss compared with that of reference software CDVS\_CPU.\fi

\begin{table}[htbp]
  \centering
  \renewcommand{\arraystretch}{1.02}
  \caption{Comparisons of CDVS CPU and GPU implementations on MPEG-CDVS database (mAP).}
    \begin{tabular}{p{0.75cm}|c|c|c|c}
     \hline
    %\toprule
          & \multicolumn{2}{c|}{Global} & \multicolumn{2}{c}{Local+Global} \\  \hline
   % \midrule
          & CPU   & GPU   & CPU   & GPU \\\hline
    512B  & 0.690   & 0.690     & 0.708    & 0.706  \\ \hline
    1K    & 0.720   & 0.720     & 0.760    & 0.760 \\ \hline
    2K    & 0.725   & 0.725     & 0.793    & 0.793  \\ \hline
    4K    & 0.780   & 0.781     & 0.820   & 0.821 \\ \hline
    8K    & 0.780   & 0.781     & 0.823   & 0.823 \\ \hline
    16K   & 0.780   & 0.781     & 0.824   & 0.823 \\ \hline
    Average & 0.746    &  0.746     &  0.788    & 0.788 \\ \hline
    %\bottomrule
    \end{tabular}%
  \label{tab:mAP_CDVS}%
\end{table}

\subsection{Speed Comparison}
In this section, we further show the speedup of the proposed CDVS encoder compared with that on CPU platform. Table \ref{tab:runningtimefeatures} shows the average running time of extracting different visual descriptors on 1000 images with resolution $640\times 480$\iffalse MPEG-CDVS database\fi. The notations CDVS\_CPU\_L and CDVS\_GPU\_L represents the running time for CDVS local feature descriptor extraction on CPU and GPU platform. Except for CDVS\_GPU and CDVS\_GPU\_L, all others descriptors are extracted on CPU platform with Intel(R) Xeon(R) CPU E5-2650 v2@2.60GHz. The CDVS\_GPU using Quadro GP100 achieves significant speedup compared with that on CPU platform when extracting $200\sim300$ local features from each image, which achieves more than 35 times \iffalse and xx times \fi speedup for CDVS\_CPU\iffalse  and OPT\_CDVS\_CPU respectively\fi. The average of CDVS extraction time is reduced from 116.69 ms on CPU platform to 3.27 ms on GPU platform. Compared with other visual feature descriptors extracted on CPU platform, the proposed CDVS encoder also significantly outperforms them, and it can well satisfy practical real-time applications.

%\begin{table}[t]
%  \centering
%  \renewcommand{\arraystretch}{1.1}
%  \caption{Comparisons of the running time for feature descriptor extraction tested on Linux PC with Intel Xeon E5-2650 v3@2.3GHz and the Tesla P100 GPU. Unit: ms }
%    \begin{tabular}{c|c|c|c}
%     \hline
%    %\toprule
%    Feature         & Time     & Feature    & Time \\ \hline
%    % \midrule
%    CDVS\_CPU       & 116.69       & SIFT  &368.88  \\  \hline
%%    OPT\_CDVS\_CPU  & 0       & SURF  &86.27  \\  \hline
%    CDVS\_GPU       & 4.17     & ORB   &12.6  \\  \hline
%    CDVS\_CPU\_L     & 102     & BRISK &19.22  \\  \hline
%%    OPT\_CDVS\_CPU\_L & 0     & AKAZE &135.55  \\  \hline
%    CDVS\_GPU\_L & 3.6       & LATCH &159.48  \\  \hline
%    \end{tabular}%
%  \label{tab:runningtimefeatures}
%\end{table}
\begin{table}[t]
  \centering
  \renewcommand{\arraystretch}{1.02}
  \caption{Comparisons of the running time for feature descriptor extraction tested on Linux PC with Intel(R) Xeon(R) CPU E5-2650 v2@2.60GHz and a Quadro GP100 GPU. Unit: ms }
    \begin{tabular}{c|c|c|c}
     \hline
    %\toprule
    Feature         & Time     &Feature    & Time \\ \hline
    % \midrule
    CDVS\_CPU       & 116.69       &SURF  &86.27  \\  \hline
    CDVS\_GPU       & 3.27     & ORB   &12.6  \\  \hline
    CDVS\_CPU\_L     & 102     & BRISK &19.22  \\  \hline
%    CDVS\_CPU\_G     & 103     & BRISK &19.22  \\  \hline
%    OPT\_CDVS\_CPU\_L & 0     & LATCH &159.48  \\  \hline
    CDVS\_GPU\_L & 3.1       & AKAZE &135.55  \\  \hline
%    CDVS\_GPU\_G & 3.08       & AKAZE &135.55  \\  \hline
    SIFT  &368.88       & LATCH &159.48  \\  \hline
    \end{tabular}%
  \label{tab:runningtimefeatures}
\end{table}

\begin{figure}[t]
\centering
%\begin{minipage}[b]{0.435\textwidth}
\includegraphics[width=6.0cm]{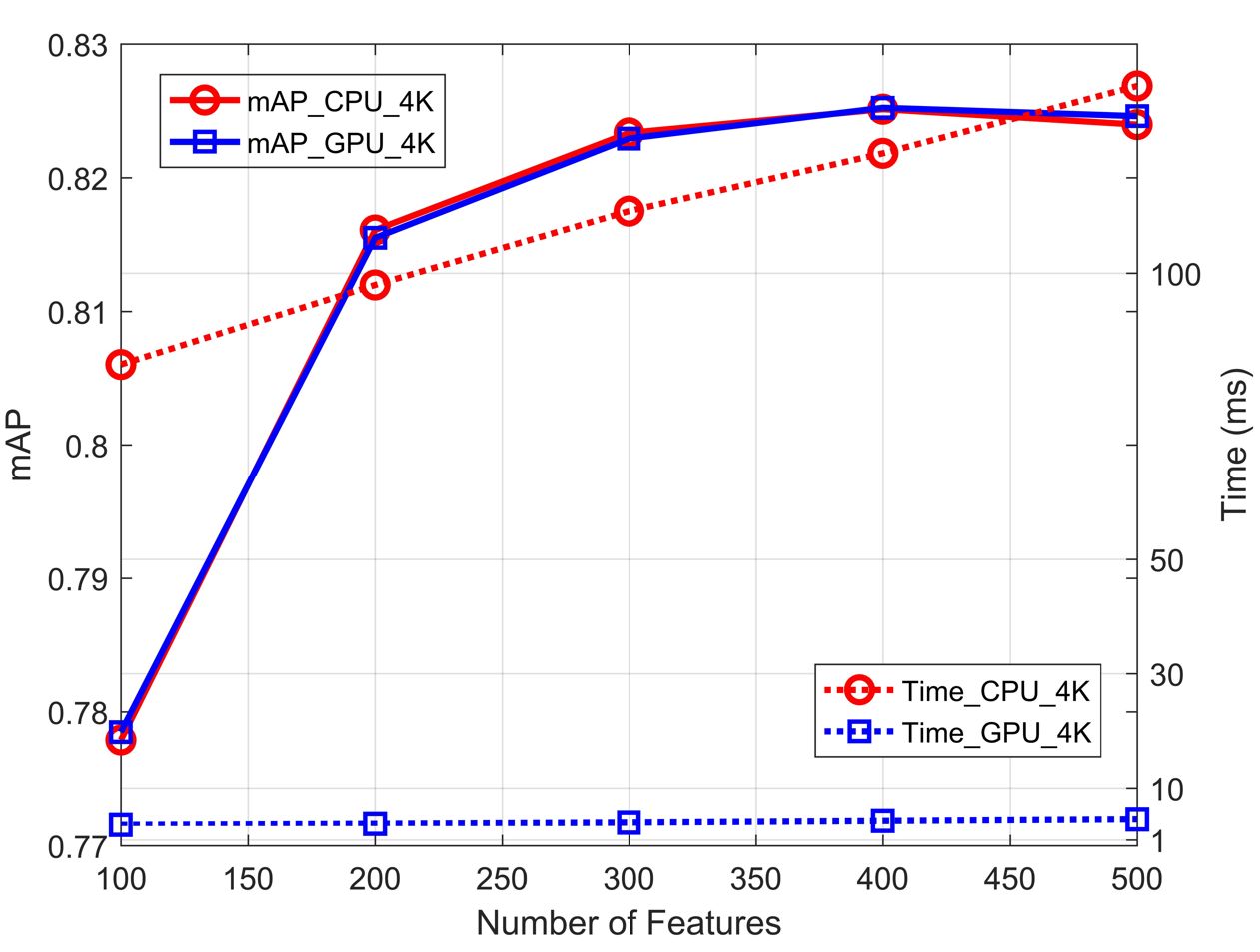}
%\subfigure[]{\label{fig:ratemodel_b}\includegraphics[width=7.0cm]{DistortionReduction.jpg} }
%\end{minipage}
\caption{The image retrieval performance vs. extraction time for CDVS on CPU and GPU platform, tested on Linux PC with Intel(R) Xeon(R) CPU E5-2650 v2@2.60GHz and a Quadro GP100 GPU. }
\label{fig:mAP_Time}
\end{figure}

Limited by computational power, the CDVS selects no more than 300 local features for each image, which may be less optimal for visual search accuracy especially for high resolution images. Hence, we further analyze the relationship between the number of local features and extraction time (Seeing Fig.~\ref{fig:mAP_Time}). We can see that with the increase of local features, the extraction time increases almost linearly on CPU. \iffalse To make a tradeoff between the computational cost and visual search accuracy, the CDVS has to select no more than 300 local features for each image.\fi However, by leveraging GPU, the extraction time cost almost keeps in the same level, about 3 ms. Therefore, more local feature descriptors can be allowed to further improve CDVS performance over GPU platforms, without noticeable increase of computational cost.

 \begin{figure}[htbp]
  \centering
 \includegraphics[width=5.0cm]{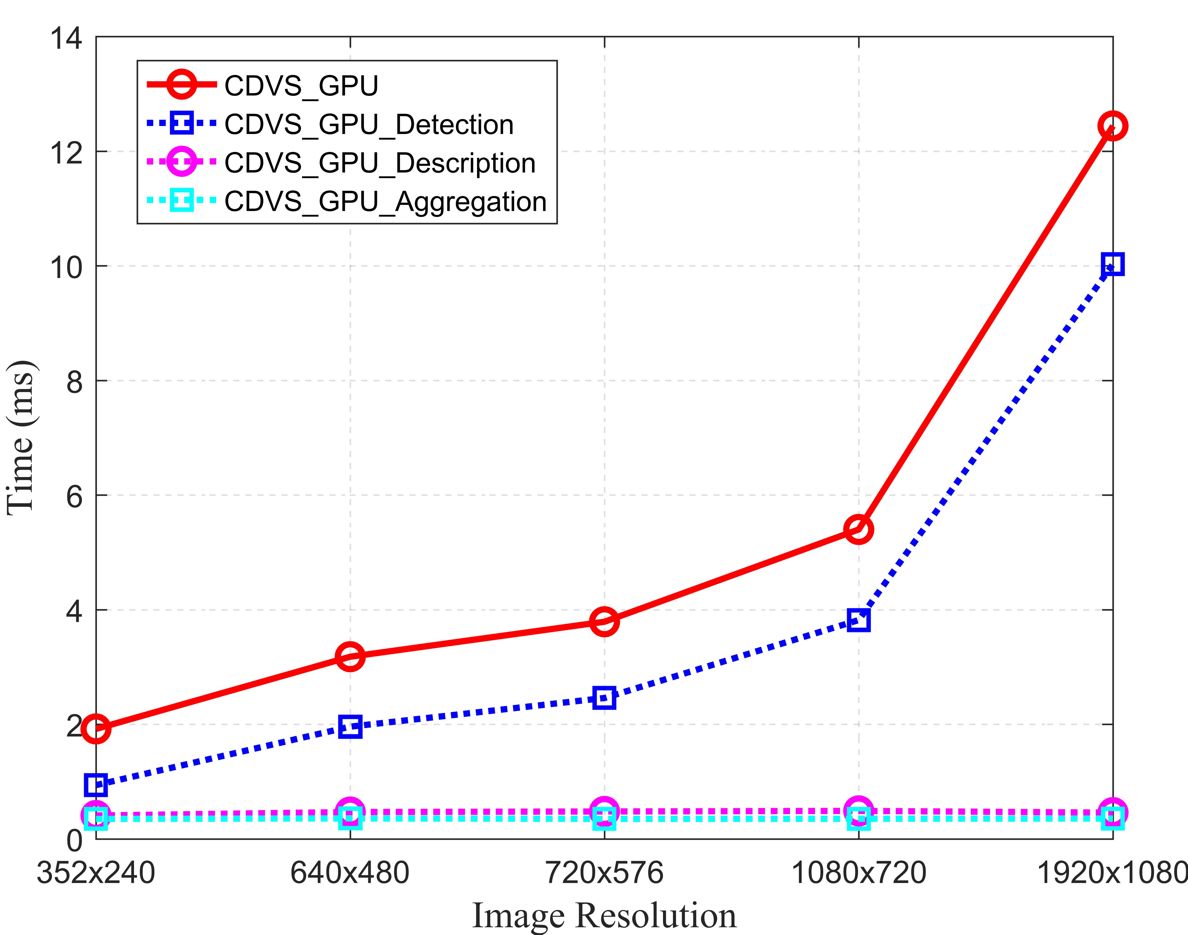}
   \caption{Influences of the resolution on the feature extraction complexity on GPU platform, tested on Linux PC with Intel(R) Xeon(R) CPU E5-2650 v2@2.60GHz and a Quadro GP100 GPU. }
\label{fig:timeresolution_gpu}
\end{figure}
 %Fig.~\ref{fig:mAP_Time} shows the relationship between the number of local features and extraction time. We can see that along with the increase of CDVS local features, the extraction time increases linearly. To make a tradeoff between the computational complexity and image retrieval performance, the CDVS usually select no more than 300 local features for each image, which is obviously not optimal for retrieval performance especially for high resolution images. However, with enough GPU resources, the extraction time for CDVS almost keeps in the same level, about 4 ms, which can well satisfy real-time applications. Therefore, more CDVS local features can be extracted to further improve the performance based on the GPU implementation.

When dealing with high resolution images, CDVS needs to first downsample the input images into low resolution with the large side no longer than 640. The downsampling operation can directly reduce the computational cost, but it also brings image distortion or information loss especially for high resolution images. We explore the variations of feature extraction time along with image resolutions for CDVS using GPU platform, and the results are shown in Fig.\ref{fig:timeresolution_gpu}. We can see that the most time-consuming module on GPU is still the interest point detection, which has been implemented on pixel-level parallelism. When the pixel number exceeds thread number, the running time will increase significantly. Therefore, the running time increases very slowly when we have enough computational resources on GPU, e.g., the image resolution smaller than $1080\times 720$. The running time for local feature description and aggregation almost does not changes since their computation mainly depend on the number of local features\iffalse which is mainly influenced by the pre-defined descriptor lengths\fi.

\begin{figure*}[t]

  \begin{minipage}[b]{0.30\textwidth}
  \centering
\subfigure[\emph{}]{\label{fig:modules_gpu_a}\includegraphics[width=5.10cm]{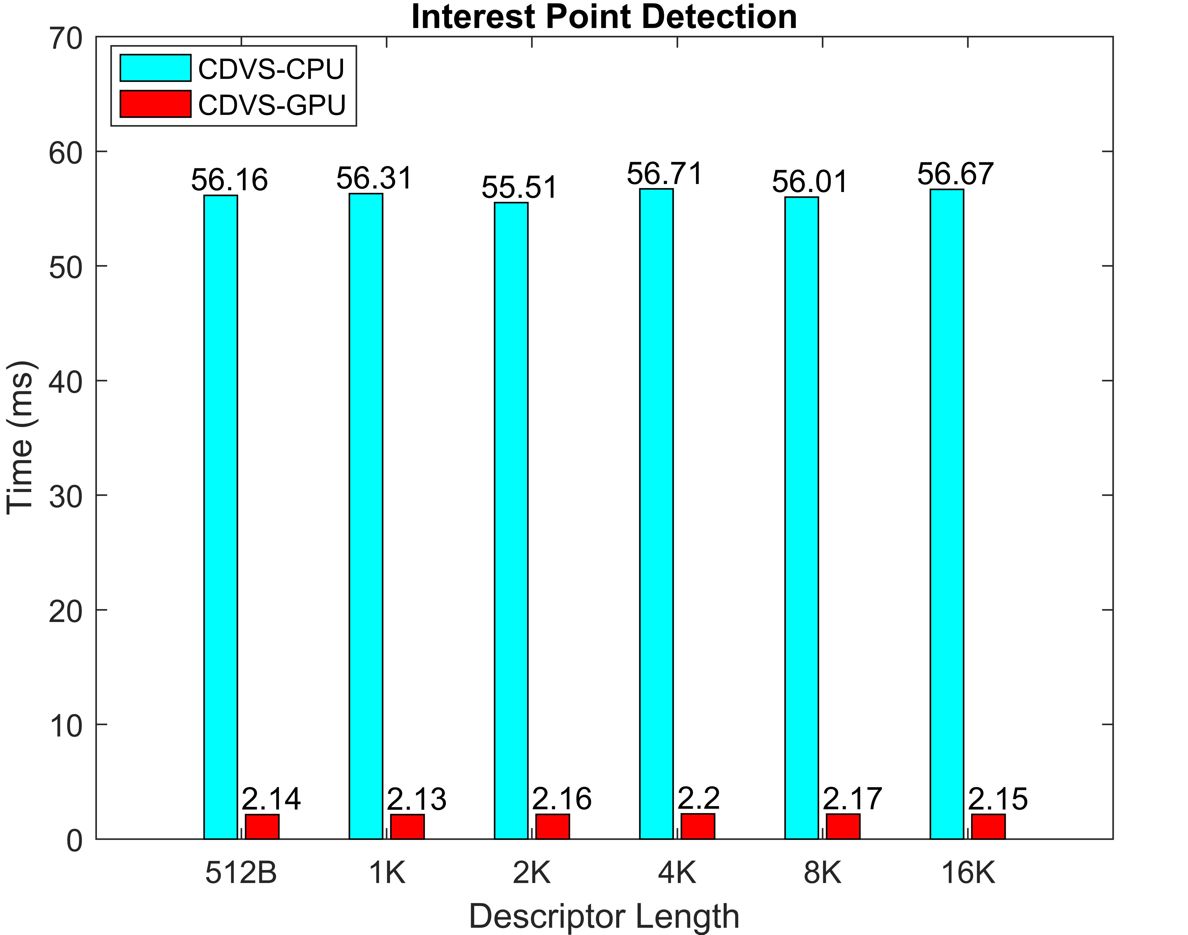} }
  \end{minipage}
  \hspace{0.1cm}
  \begin{minipage}[b]{0.30\textwidth}
  \centering
  \subfigure[\emph{}]{\label{fig:modules_gpu_b}\includegraphics[width=5.10cm]{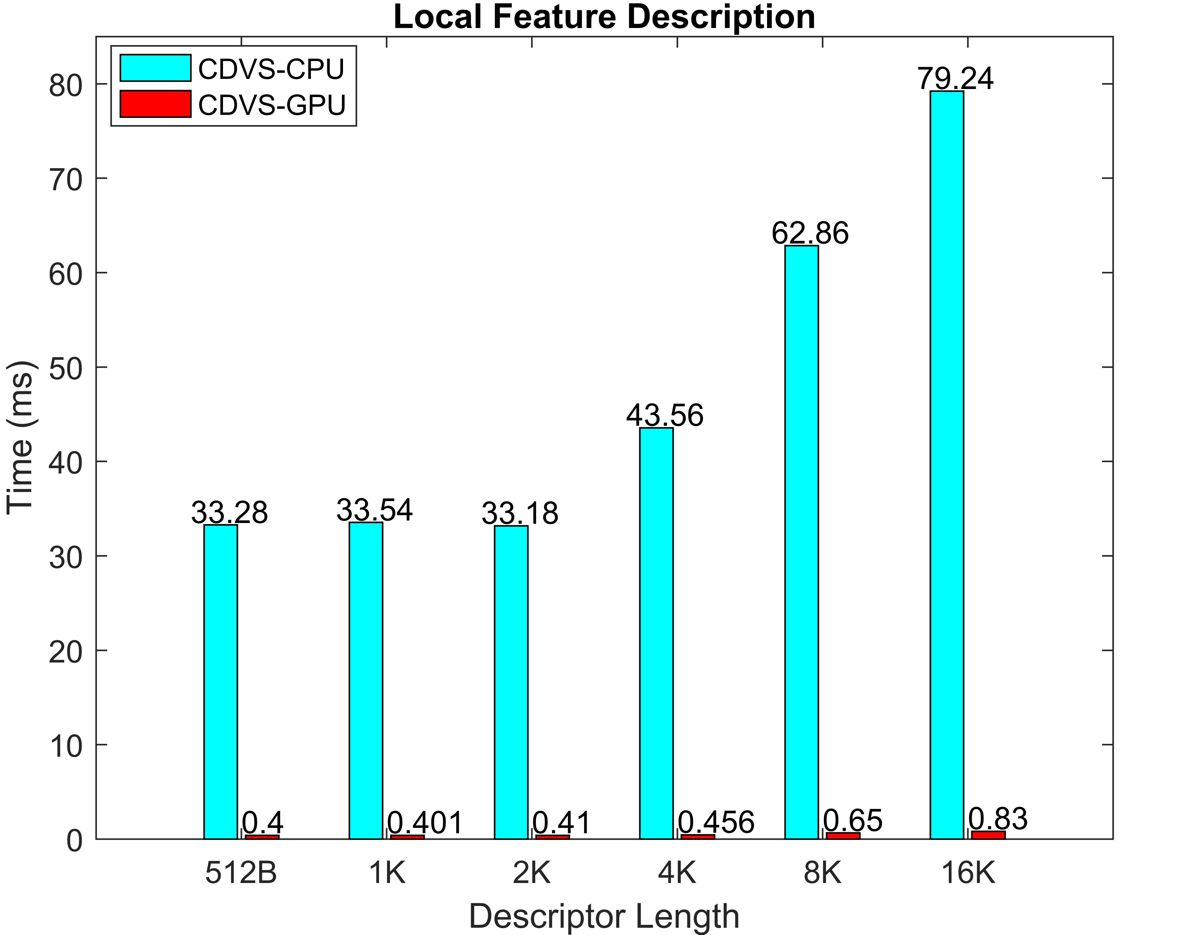} }
  \end{minipage}
  \hspace{0.1cm}
  \begin{minipage}[b]{0.30\textwidth}
  \centering
  \subfigure[\emph{}]{\label{fig:modules_gpu_c}\includegraphics[width=5.10cm]{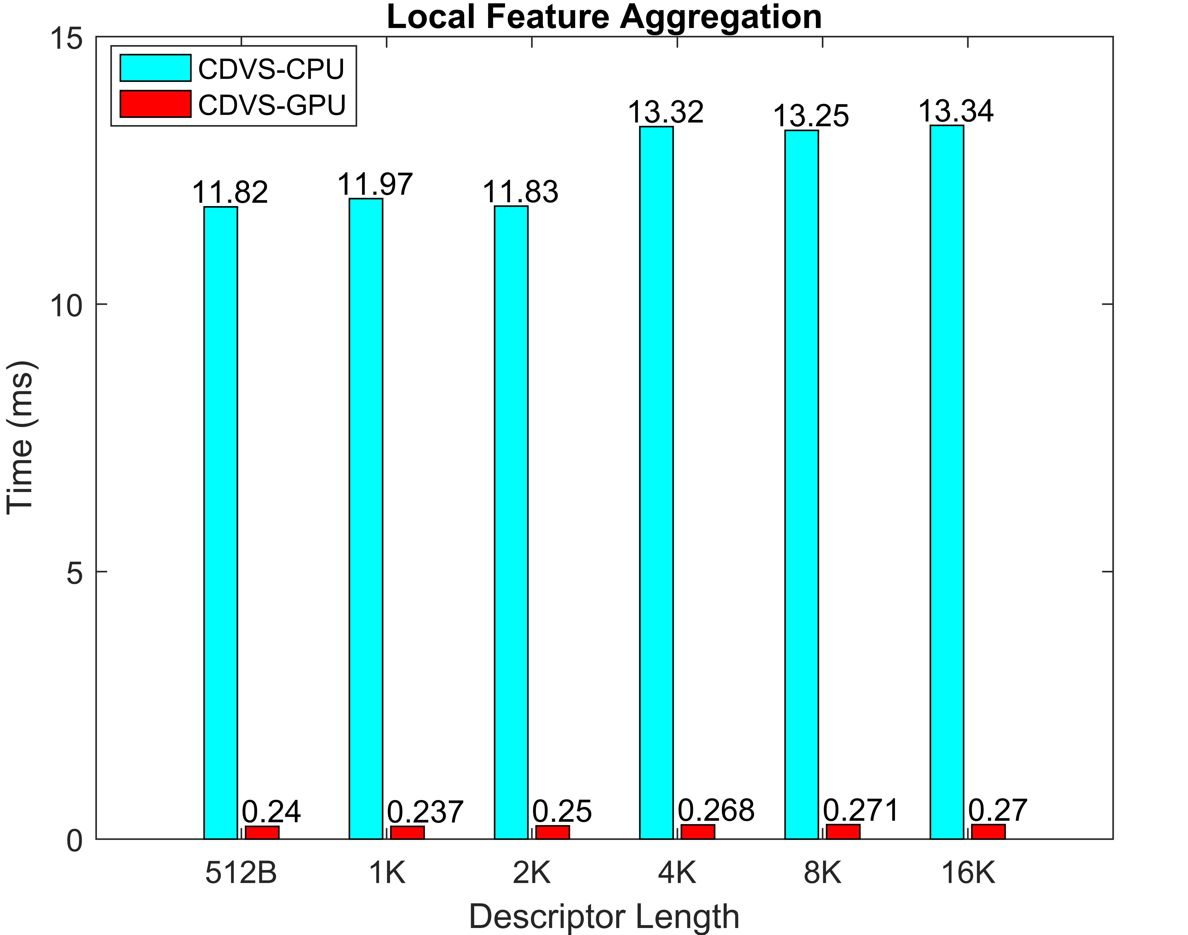} }
  \end{minipage}
  \caption{Complexity comparisons for different modules. (a) Interest point detection; (b) Local feature description; (c) Local feature aggregation. These results are tested on on Linux PC with Intel(R) Xeon(R) CPU E5-2650 v2@2.60GHz and a Quadro GP100 GPU.}
\label{fig:modules_gpu_time}
\end{figure*}

\begin{figure}[t]
\centering
%\begin{minipage}[b]{0.435\textwidth}
\includegraphics[width=7.0cm]{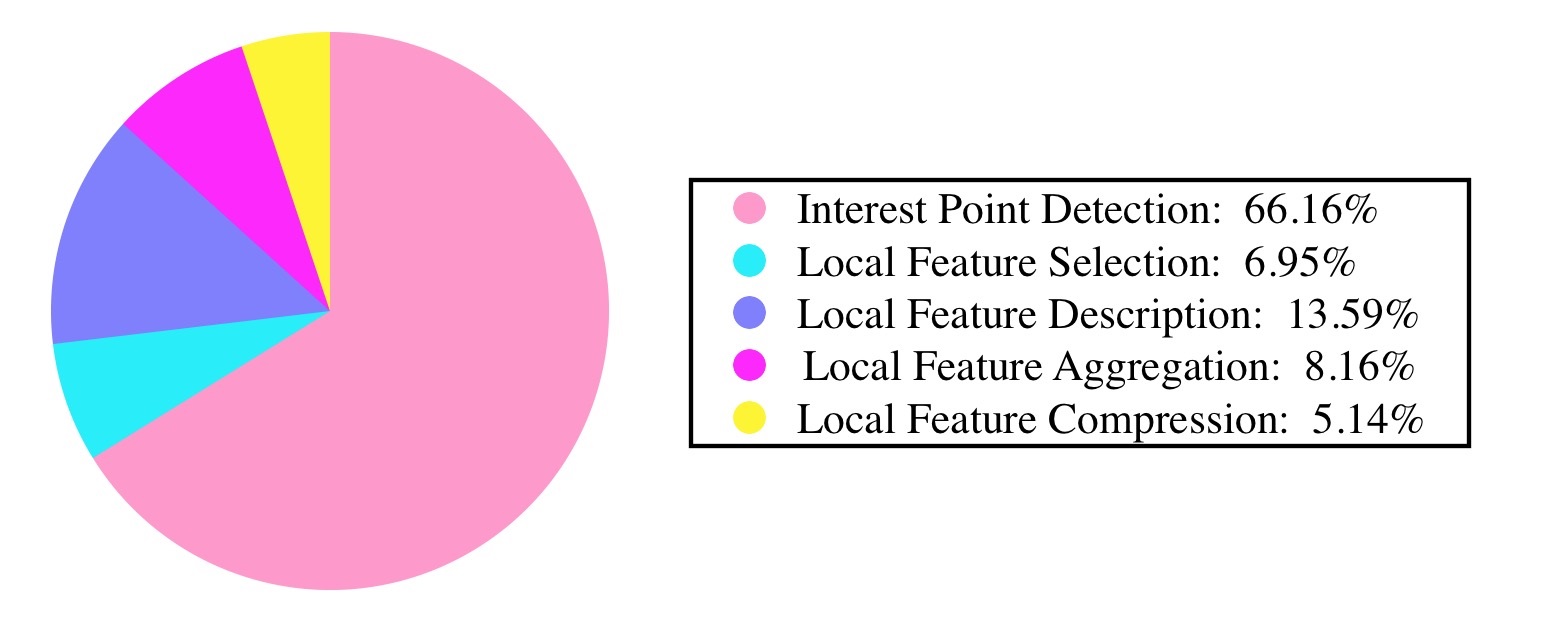}
%\subfigure[]{\label{fig:ratemodel_b}\includegraphics[width=7.0cm]{./DistortionReduction.jpg} }
%\end{minipage}
\caption{The running time consumption for different modules of the proposed fast CDVS, tested on Linux PC with Intel(R) Xeon(R) CPU E5-2650 v2@2.60GHz and a Quadro GP100 GPU. }
\label{fig:TimeDistributionGPU}
\end{figure}

To explore the speedup on GPU, we test the running time for these modules on CPU and GPU respectively, and Fig.~\ref{fig:modules_gpu_time} shows their running time at different pre-defined descriptor length. The CDVS\_GPU achieves about 26, 50 and 22 times speedup for interest point detection, local feature description and aggregation compared with that of CDVS\_CPU, respectively. Fig.\ref{fig:TimeDistributionGPU} shows the running time proportion for different modules with GPU-CPU hybrid computing. We can see that the running time consumption can be covered by the local feature aggregation when using GPU-CPU hybrid computing.

To further test the speedup of the fast CDVS encoder, we run tests on different kinds of GPUs as shown in Table \ref{tab:runningtimeGPUs}. Herein, the 11 NVIDIA GPUs are utilized in our experiment, Tesla M40, GTX1060, GTX1080 \iffalse,  GTX1060\fi and GTX1080Ti are the popular GPUs targeting at hyper-converged systems, and the Jetson TX1 is the popular GPU for embedded systems. Specially, NVIDIA has provided us the up-to-date GPUs for the two systems, i.e., Tesla P40, P100, Titan X, Quadro GP100 and Jetson TX2, which shows the maximum speedup. Our CDVS encoder only costs $3.38\sim 5.05$ ms for VGA resolution images on popular GPUs, and $3.24\sim 3.75$ ms on the up-to-date GPUs. Even for the $1920\times 1080$ images, our encoder can extract the descriptors in real time with minimum 12.41 ms. More importantly, our CDVS encoder can extract feature descriptors for $640\times 480$ images in real time on Jetson TX2, which are the promising platform for embedded systems with low power consumption. This indicates that the CDVS can be well deployed on mobile devices to support very fast visual search. Based on comprehensive experimental results, we can claim that the CDVS using GPU-CPU hybrid computing can well support scalable image/video retrieval and analysis with real-time requirements.

\renewcommand{\thefootnote}{\fnsymbol{footnote}}
\renewcommand\arraystretch{0.2}
\begin{table}[htbp]
  \centering
  \renewcommand{\arraystretch}{1.02}
  \caption{The running time comparison of CDVS on different GPU platform, tested on Linux PC with Intel(R) Xeon(R) CPU E5-2650 v2 @ 2.60GHz, Time (ms). }
    \begin{tabular}{m{1.5cm}<{\centering}|p{0.7cm}<{\centering}|p{0.7cm}<{\centering}|m{1.80cm}|p{0.7cm}<{\centering}|p{0.7cm}<{\centering}}
     \hline
   % \toprule
    GPUs & $640\times480$ & $1920\times1080$  &GPUs  & $640\times 480$ & $1920\times 1080$\\ \hline
    %\midrule
    GTX 1060                       &5.05   &20.91      &GTX 1080         &3.54   &12.97   \\ \hline
    Tesla M40                      &5.02   &20.4       &GTX 1080Ti    &3.38   &12.58\\ \hline
    Titan X Pascal$^*$             &3.75    &15.37     &Quadro GP100$^*$     &3.27   &12.41\\ \hline
    Tesla P40\tablefootnote{The state-of-the-art GPU provided by NVIDIA}            &3.7    &14.38   &Jetson TX1     &40.22    &161.51       \\ \hline
    Tesla P100$^*$   &3.56    &14.61                   &Jetson TX2$^*$     &28.69    &138.74                                            \\ \hline
  %  \bottomrule
    \end{tabular}%
  \label{tab:runningtimeGPUs}%
\end{table}%

\begin{table*}[htbp]
  \centering
  \renewcommand{\arraystretch}{1.02}
  \caption{Performance comparison of CNNs, CDVS and their combination descriptors, tested on Linux PC with Intel(R) Xeon(R) CPU E5-2650 v2@2.60GHz and a Quadro GP100 GPU. }
    \begin{tabular}{p{3.5cm}|c|c|c|c|c|c}
     \hline
   % \toprule
                  & mAP & Top Match  &TPR@FPR=0.01 & Descriptor Size & Extraction time (ms) & Memory\\ \hline
    %\midrule
    CDVS                       &0.8203   &0.9093  &0.918    &4 KB   &3.27    &205 MB \\ \hline
    RMAC                       &0.57    &0.6638   &0.456    &2 KB   &36.5   &6635 MB \\ \hline
    RMAC + CDVS local          &0.7858   &0.8921  &0.913    &4 KB   &40.1   &6715 MB  \\ \hline
    RMAC + CDVS global         &0.8009   &0.872   &0.8714   &4 KB    &40.2   &6840 MB \\ \hline
    NIP-VGG-16                 &0.6768   &0.7809  &0.52     &2 KB   &144    &6635 MB \\ \hline
    NIP-VGG-16 + CDVS local    &0.8003   &0.865   &0.9166   &4 KB   &147.6  &6715 MB \\ \hline
    NIP-VGG-16 + CDVS global   &0.8508   &0.9124  &0.874    &4 KB   &147.7  &6840 MB \\ \hline
  %  \bottomrule
    \end{tabular}%
  \label{tab:NIPCNN_CDVS}%
\end{table*}%

\subsection{Promising Future of CDVS and CNN Feature Descriptors}
CDVS provides MPEG standard compliant handcrafted visual feature descriptor with high performance, good compactness and friendly to parallel implementation. Recently, although the visual feature descriptors generated from convolutional neural network (CNN) have provide very promising performance, the performance can be further improved when combining the handcraft visual descriptors of CDVS. In \cite{Lou_DCC2017}, we have proposed a Nested Invariance Pooling (NIP) method  to obtain compact and robust CNN descriptors, which are generated by applying three different pooling operations to the feature maps of CNNs in a nested way towards rotation and scale invariant feature representation. To explore the potential performance, we combine the CDVS local and global feature descriptors with two CNN features, i.e., NIP-VGG-16 \cite{Lou_DCC2017} and RMAC \cite{tolias2015particular} to perform pair matching and image retrieval.

In our experiments, the dimensions of NIP-VGG-16 and RMAC descriptors are 512. We use 4 bytes to represent each dimension, which leads to 2KB representation for NIP-VGG-16 and RMAC. Table \ref{tab:NIPCNN_CDVS} shows the image retrieval and matching performance on MPEG-CDVS database. \iffalse Another metric, Precision@R, is adopted in this experiment, which is the retrieval precision at a given cut-off rank R for a single query.\fi Although both CNN feature descriptors achieve good performance at low bitrate, their performances are further improved by combining CDVS descriptors. The improvements of NIP-VGG-16 with CDVS global descriptors are up to 0.0305 and 0.174 compared with CDVS and NIP-VGG-16 respectively in terms of mAP. \iffalse With FPR=1\%, TPR gets more than xx\% and xx\% improvements.\fi To the best of our knowledge, the combination of CNN and CDVS achieves the best pair matching and image retrieval performance at the same descriptor length, which also has been verified in the latest proposals of the emerging MPEG Compact Descriptors for Video Analysis (CDVA)standard \cite{ISO2012_MPEGCDVS39219}.

% In our experiments, the two CNN model are utilized to generate the NIP CNN features, i.e., VGG-16 \cite{simonyan2014very} and AlexNet \cite{NIPS2012_4824} networks, and the dimensions of NIP CNNs descriptors of VGG-16 and AlexNet are 512 and 256 respectively. We utilize 4 bytes to represent each dimension, which leads to 1KB representation of NIP CNNs and 2KB for AlexNet and VGG-16 respectively. Table \ref{tab:NIPCNN_CDVS} shows the image retrieval and matching performance on MPEG-CDVS database. Another metric, Precision@R, is adopted in this experiment, which is the retrieval precision at a given cut-off rank R for a single query. Although both of the two CNN features achieve good performance at low bitrate, their performances are further improved significantly by combining CDVS features. The improvements of NIP CNN and CDVS global descriptors are up to xx\% and xx\% compared with CDVS and NIP CNN respectively based on mAP. With FPR=1\%, TPR get more than xx\% and xx\% improvements. To the best of our knowledge, the combination of CNN and CDVS achieves the best image retrieval and matching results at the same bitrate scenario.

Although the CNN descriptors achieve promising results in computer vision applications, the extremely huge computational burdens make them heavily dependent on GPU platform. For 1000 VGA resolution images, the average running time of CNN descriptor extraction about 144 ms and 36.5 ms for NIP-VGG-16 and RMAC networks, which obviously exceed that of CDVS on the GPU platform. In additional, the CNN descriptor extraction also consumes too much memory compared with that of CDVS as shown in Table \ref{tab:NIPCNN_CDVS}. Hence,it is promising to combine the CNN feature descriptors and the handcraft feature descriptors, and implement them harmonically using GPU platform, which can provide scalable descriptors for both computational resources and visual search accuracy.

%Although the CNN descriptors achieve promising results in most of computer vision applications, the huge computational burdens make them heavily dependent on GPU platform. On MPEG-CDVS database, the average running time of CNN descriptor extraction about xx ms and xx ms for VGG-16 and AlexNet networks, which obviously exceed that of CDVS on the same GPU platform. In additional, we also compared the memory and bandwidth consume for both CDVS and CNN descriptors. (Insert a paragraph of memory and bandwidth analysis to show that the CDVS are more efficient in GPU resources). Finally, the CDVS also can be well performed on GPU platform, which show a good research...

\section{Conclusion}
We have revisited the merits of the MPEG-CDVS standard in computational cost reduction. A very fast CDVS encoder has been implemented using hybrid GPU-CPU computing. By thoroughly comparison with other state-of-the-art visual descriptors on large-scale database, the fast CDVS encoder achieves significant speedup compared with that on CPU platform (more than 35 times) while maintaining the competitive performance for image retrieval and matching. Furthermore, by incorporating the CDVS encoder with deep learning based approaches on GPU platform, we have shown that the handcraft visual feature descriptors and CNN based feature descriptors are complementary to some extent and the combination of CDVS descriptors and the CNN descriptors has achieved the state-of-the-art visual search performance over benchmarks. Especially towards real-time (mobile) visual search and augmented reality applications, how to harmoniously leverage the merits of highly efficient and low complexity handcrafted descriptors, and the state-of-the-art CNNs based descriptors via GPU or GPU-CPU hybrid computing, is a competitive and promising topic.

%We have presented an efficient GPU based implementation strategy for MPEG CDVS standard. By thoroughly analyzing the complexity of CDVS, we identify the modules that are appropriate for GPU implementation and design a practical CDVS implementation strategy, which is fully optimized in terms of the complexity-accuracy performance. Extensive experimental results show that the proposed scheme can achieve xxx ms/pic while maintaining the image retrieval performance, which is also practical for real-time CDVS based video feature extraction. Further incorporation of the proposed scheme with deep learning based approaches demonstrates that the extraction of CDVS and deep learning features can be unified on GPU platform to achieve the state-of-the-art visual search performance.

% or


\ifCLASSOPTIONcaptionsoff
  \newpage
\fi



%\bibliographystyle{IEEEtran}
%\bibliography{IEEEabrv,cdvs}

\begin{thebibliography}{10}
\providecommand{\url}[1]{#1}
\csname url@samestyle\endcsname
\providecommand{\newblock}{\relax}
\providecommand{\bibinfo}[2]{#2}
\providecommand{\BIBentrySTDinterwordspacing}{\spaceskip=0pt\relax}
\providecommand{\BIBentryALTinterwordstretchfactor}{4}
\providecommand{\BIBentryALTinterwordspacing}{\spaceskip=\fontdimen2\font plus
\BIBentryALTinterwordstretchfactor\fontdimen3\font minus
  \fontdimen4\font\relax}
\providecommand{\BIBforeignlanguage}[2]{{%
\expandafter\ifx\csname l@#1\endcsname\relax
\typeout{** WARNING: IEEEtran.bst: No hyphenation pattern has been}%
\typeout{** loaded for the language `#1'. Using the pattern for}%
\typeout{** the default language instead.}%
\else
\language=\csname l@#1\endcsname
\fi
#2}}
\providecommand{\BIBdecl}{\relax}
\BIBdecl

\bibitem{girod2011mobile1}
B.~Girod, V.~Chandrasekhar, R.~Grzeszczuk, and Y.~A. Reznik, ``Mobile visual
  search: Architectures, technologies, and the emerging mpeg standard,''
  \emph{IEEE MultiMedia}, vol.~18, no.~3, pp. 86--94, 2011.

\bibitem{lowe2004distinctive}
D.~G. Lowe, ``Distinctive image features from scale-invariant keypoints,''
  \emph{International journal of computer vision}, vol.~60, no.~2, pp. 91--110,
  2004.

\bibitem{bay2008speeded}
H.~Bay, A.~Ess, T.~Tuytelaars, and L.~Van~Gool, ``Speeded-up robust features
  (surf),'' \emph{Computer vision and image understanding}, vol. 110, no.~3,
  pp. 346--359, 2008.

\bibitem{Rublee_ORB2011}
E.~Rublee, V.~Rabaud, K.~Konolige, and G.~Bradski, ``Orb: An efficient
  alternative to sift or surf,'' in \emph{2011 International Conference on
  Computer Vision}, Nov 2011, pp. 2564--2571.

\bibitem{Leutenegger_2011}
S.~Leutenegger, M.~Chli, and R.~Y. Siegwart, ``Brisk: Binary robust invariant
  scalable keypoints,'' in \emph{2011 International Conference on Computer
  Vision}, Nov 2011, pp. 2548--2555.

\bibitem{duan2016overview}
L.-Y. Duan, V.~Chandrasekhar, J.~Chen, J.~Lin, Z.~Wang, T.~Huang, B.~Girod, and
  W.~Gao, ``Overview of the {MPEG-CDVS} standard,'' \emph{IEEE Transactions on
  Image Processing}, vol.~25, no.~1, pp. 179--194, 2016.

\bibitem{mpeg_cdvs_standard}
``{Information technology-Multimedia content description interface-Part 13:
  Compact descriptors for visual search},'' \emph{ISO/IEC
  JTC1/SC29/WG11/N14956}, Oct 2014.

\bibitem{Chen_BFLoG2015}
J.~Chen, L.~Y. Duan, F.~Gao, J.~Cai, A.~C. Kot, and T.~Huang, ``A low
  complexity interest point detector,'' \emph{IEEE Signal Processing Letters},
  vol.~22, no.~2, pp. 172--176, Feb 2015.

\bibitem{garland2008parallel}
M.~Garland, S.~Le~Grand, J.~Nickolls, J.~Anderson, J.~Hardwick, S.~Morton,
  E.~Phillips, Y.~Zhang, and V.~Volkov, ``Parallel computing experiences with
  cuda,'' \emph{Micro, IEEE}, vol.~28, no.~4, pp. 13--27, 2008.

\bibitem{clara2008nvidia}
S.~Clara, ``Nvidia cuda compute unified device architecture: Programming guide
  version 1.1.''

\bibitem{heymann2007sift}
S.~Heymann, K.~M{\"u}ller, A.~Smolic, B.~Froehlich, and T.~Wiegand, ``{SIFT
  implementation and optimization for general-purpose GPU},''
  \emph{International Conference in Central Europe on Computer Graphics,
  Visualization and Computer Vision}, 2007.

\bibitem{wu2007siftgpu}
C.~Wu, ``{SiftGPU: A GPU implementation of scale invariant feature transform
  (SIFT)},'' 2007.

\bibitem{rister2013fast}
B.~Rister, G.~Wang, M.~Wu, and J.~R. Cavallaro, ``{A fast and efficient SIFT
  detector using the mobile GPU},'' in \emph{2013 IEEE International Conference
  on Acoustics, Speech and Signal Processing}, 2013, pp. 2674--2678.

\bibitem{lee2016complexity}
C.~Lee, C.~E. Rhee, and H.-J. Lee, ``{Complexity Reduction by Modified
  Scale-Space Construction in SIFT Generation Optimized for a Mobile GPU},''
  \emph{IEEE Transactions on Circuits and Systems for Video Technology}, 2016.

\bibitem{cudasift}
CUDASIFT, ``{https://github.com/Celebrandil/CudaSift},'' 2007.

\bibitem{wang2013workload}
G.~Wang, B.~Rister, and J.~R. Cavallaro, ``{Workload analysis and efficient
  OpenCL-based implementation of SIFT algorithm on a smartphone},'' in
  \emph{Global Conference on Signal and Information Processing (GlobalSIP),
  2013 IEEE}, 2013, pp. 759--762.

\bibitem{patlolla2015gpu}
D.~Patlolla, S.~Voisin, H.~Sridharan, and A.~Cheriyadat, ``{GPU accelerated
  textons and dense sift features for human settlement detection from
  high-resolution satellite imagery},'' \emph{GeoComp}, 2015.

\bibitem{cornelis2008fast}
N.~Cornelis and L.~Van~Gool, ``Fast scale invariant feature detection and
  matching on programmable graphics hardware,'' in \emph{Computer Vision and
  Pattern Recognition Workshops, 2008. CVPRW'08. IEEE Computer Society
  Conference on}, 2008, pp. 1--8.

\bibitem{ma2016gpu}
W.~Ma, L.~Cao, L.~Yu, G.~Long, and Y.~Li, ``{GPU-FV: Realtime Fisher Vector and
  Its Applications in Video Monitoring},'' \emph{arXiv preprint
  arXiv:1604.03498}, 2016.

\bibitem{krizhevsky2012imagenet}
A.~Krizhevsky, I.~Sutskever, and G.~E. Hinton, ``Imagenet classification with
  deep convolutional neural networks,'' in \emph{Advances in neural information
  processing systems}, 2012, pp. 1097--1105.

\bibitem{zheng2016sift}
L.~Zheng, Y.~Yang, and Q.~Tian, ``{SIFT meets CNN: a decade survey of instance
  retrieval},'' \emph{arXiv preprint arXiv:1608.01807}, 2016.

\bibitem{jia2014caffe}
Y.~Jia, E.~Shelhamer, J.~Donahue, S.~Karayev, J.~Long, R.~Girshick,
  S.~Guadarrama, and T.~Darrell, ``Caffe: Convolutional architecture for fast
  feature embedding,'' in \emph{Proceedings of the 22nd ACM international
  conference on Multimedia}.\hskip 1em plus 0.5em minus 0.4em\relax ACM, 2014,
  pp. 675--678.

\bibitem{abadi2016tensorflow}
M.~Abadi, A.~Agarwal, P.~Barham, E.~Brevdo, Z.~Chen, C.~Citro, G.~S. Corrado,
  A.~Davis, J.~Dean, M.~Devin \emph{et~al.}, ``Tensorflow: Large-scale machine
  learning on heterogeneous distributed systems,'' \emph{arXiv preprint
  arXiv:1603.04467}, 2016.

\bibitem{ISO2013_MPEGCDVS28076}
D.~Pau, E.~Napoli, G.~Lopez, E.~Plebani, A.~BRUNA, and D.~SORENSEN, ``{Fourier
  transform Based interest point detector using LoG frequency response},''
  \emph{{ISO/IEC JTC1/SC29/WG11/M28076}}, Jan 2013.

\bibitem{ISO2013_MPEGCDVS28090}
Z.~Liu, Q.~Zhou, and X.~Guojun, ``{Huawei's Response to CE 4: Preliminary
  Results by Fourier Transform Based LOG},'' \emph{{ISO/IEC
  JTC1/SC29/WG11/M28090}}, Jan 2013.

\bibitem{ISO2014_MPEGCDVS33159}
C.~Jie, L.-Y. Duan, T.~Huang, W.~Gao, A.~C. Kot, M.~Balestri, G.~Francini, and
  S.~Leps{\o}y, ``{CDVS CE1: A low complexity detector ALP\_BFLoG},''
  \emph{{ISO/IEC JTC1/SC29/WG11/M33159}}, Oct 2014.

\bibitem{ISO2013_MPEGCDVS31399}
C.~Jie, L.-Y. Duan, T.~Huang, and W.~Gao, ``{Peking University Response to CE1:
  Improved BFLoG Interest Point Detector},'' \emph{{ISO/IEC
  JTC1/SC29/WG11/M31399}}, Oct 2013.

\bibitem{ISO2013_MPEGCDVS30256}
G.~Francini, S.~Lepsoy, and M.~Balestri, ``{CDVS: Telecom Italia's response to
  CE1 – interest point detection},'' \emph{{ISO/IEC JTC1/SC29/WG11/M30256}},
  Jul 2013.

\bibitem{carr1999option}
P.~Carr and D.~Madan, ``Option valuation using the fast fourier transform,''
  \emph{Journal of computational finance}, vol.~2, no.~4, pp. 61--73, 1999.

\bibitem{ISO2012_MPEGCDVS23822}
L.-T. Cheok, J.~Song, and K.~Park, ``{CDVS: Telecom Italia's response to CE1 –
  interest point detection},'' \emph{{ISO/IEC JTC1/SC29/WG11/M23822}}, Feb
  2012.

\bibitem{ISO2012_MPEGCDVS23929}
W.~Chunyu, L.-Y. Duan, C.~Jie, T.~Huang, and W.~Gao, ``{Reference results of
  key point reduction},'' \emph{ISO/IEC JTC1/SC29/WG11/M23929}, Feb 2012.

\bibitem{francini2013selection}
G.~Francini, S.~Leps{\o}y, and M.~Balestri, ``Selection of local features for
  visual search,'' \emph{Signal Processing: Image Communication}, vol.~28,
  no.~4, pp. 311--322, 2013.

\bibitem{ISO2012_MPEGCDVS12367}
G.~Francini, S.~Lepsoy, and M.~Balestri, ``{Description of Test Model under
  Consideration for CDVS},'' \emph{{ISO/IEC JTC1/SC29/WG11/N12367}}, Feb 2012.

\bibitem{ISO2012_MPEGCDVS24737}
------, ``{Telecom Italia Response to the CDVS Core Experiment 2},''
  \emph{{ISO/IEC JTC1/SC29/WG11/M24737}}, Apr 2012.

\bibitem{jegou2011product}
H.~Jegou, M.~Douze, and C.~Schmid, ``Product quantization for nearest neighbor
  search,'' \emph{Pattern Analysis and Machine Intelligence, IEEE Transactions
  on}, vol.~33, no.~1, pp. 117--128, 2011.

\bibitem{ISO2012_MPEGCDVS22806}
J.~Chen, L.-Y. Duan, C.~Wang, T.~Huang, and W.~Gao, ``{Peking Univ. Response to
  CE 2: Improvements of the SCFV Global Descriptor},'' \emph{{ISO/IEC
  JTC1/SC29/WG11/M22806}}, Oct 2011.

\bibitem{ISO2012_MPEGCDVS24780}
C.~Jie, L.-Y. Duan, T.~Huang, and W.~Gao, ``{CDVS:CE2: Multi-Stage Vector
  Quantization for Low Memory Descriptors},'' \emph{{ISO/IEC
  JTC1/SC29/WG11/M24780}}, Apr 2012.

\bibitem{chen2011residual}
D.~Chen, S.~Tsai, V.~Chandrasekhar, G.~Takacs, H.~Chen, R.~Vedantham,
  R.~Grzeszczuk, and B.~Girod, ``Residual enhanced visual vectors for on-device
  image matching,'' in \emph{Signals, Systems and Computers (ASILOMAR), 2011
  Conference Record of the Forty Fifth Asilomar Conference on}.\hskip 1em plus
  0.5em minus 0.4em\relax IEEE, 2011, pp. 850--854.

\bibitem{ISO2012_MPEGCDVS23578}
D.~Chen, V.~Chandrasekhar, G.~Takacs, S.~Tsai, M.~Makar, R.~Vedantham,
  R.~Grzeszczuk, and B.~Girod, ``{Improvements to the Test Model Under
  Consideration with a Global Descriptor},'' \emph{{ISO/IEC
  JTC1/SC29/WG11/M23578}}, Feb 2012.

\bibitem{husain2016improving}
S.~S. Husain and M.~Bober, ``Improving large-scale image retrieval through
  robust aggregation of local descriptors,'' \emph{IEEE Transactions on Pattern
  Analysis and Machine Intelligence}, 2016.

\bibitem{ISO2012_MPEGCDVS31426}
M.~Bober, S.~Husain, S.~Paschalakis, and K.~Wnukowicz, ``{Improving performance
  and usability of CDVS TM7 with a Robust Visual Descriptor (RVD) - CE 2
  Proposal from University of Surrey and Visual Atoms},'' \emph{{ISO/IEC
  JTC1/SC29/WG11/M31426}}, Oct 2013.

\bibitem{lin2014rate}
J.~Lin, L.-Y. Duan, Y.~Huang, S.~Luo, T.~Huang, and W.~Gao, ``Rate-adaptive
  compact fisher codes for mobile visual search,'' \emph{IEEE Signal Processing
  Letters}, vol.~21, no.~2, pp. 195--198, 2014.

\bibitem{ISO2012_MPEGCDVS31401}
J.~Lin, L.-Y. Duan, Z.~Wang, T.~Huang, and W.~Gao, ``{Peking Univ. Response to
  CE 2: Improvements of the SCFV Global Descriptor},'' \emph{{ISO/IEC
  JTC1/SC29/WG11/M31401}}, Oct 2013.

\bibitem{ISO2011_MPEGCDVS}
``Evaluation framework for compact descriptors for visual search,''
  \emph{{ISO/IEC JTC1/SC29/WG11/N12202}}, Jul 2011.

\bibitem{jegou2008hamming}
H.~Jegou, M.~Douze, and C.~Schmid, ``Hamming embedding and weak geometric
  consistency for large scale image search,'' \emph{Computer Vision--ECCV
  2008}, pp. 304--317, 2008.

\bibitem{Vijay2013_PatchCDVS}
\BIBentryALTinterwordspacing
{CDVS Patches}, 2013. [Online]. Available:
  \url{http://blackhole1.stanford.edu/vijayc/cdvs patches.tar}
\BIBentrySTDinterwordspacing

\bibitem{winder2009picking}
S.~Winder, G.~Hua, and M.~Brown, ``Picking the best daisy,'' in \emph{Computer
  Vision and Pattern Recognition, 2009. CVPR 2009. IEEE Conference on}.\hskip
  1em plus 0.5em minus 0.4em\relax IEEE, 2009, pp. 178--185.

\bibitem{Alcantarilla2013Fast}
P.~Alcantarilla, J.~Nuevo, and A.~Bartoli, ``Fast explicit diffusion for
  accelerated features in nonlinear scale spaces,'' in \emph{British Machine
  Vision Conference}, 2013, pp. 13.1--13.11.

\bibitem{Levi_2016}
G.~Levi and T.~Hassner, ``Latch: Learned arrangements of three patch codes,''
  in \emph{2016 IEEE Winter Conference on Applications of Computer Vision
  (WACV)}, March 2016, pp. 1--9.

\bibitem{Lou_DCC2017}
Y.~Lou, Y.~Bai, J.~Lin, S.~Wang, J.~Chen, V.~Chandrasekhar, L.-Y. Duan,
  T.~Huang, A.~C. Kot, and W.~Gao, ``Compact deep invariant descriptors for
  video retrieval,'' in \emph{2017 Data Compression Conference}, 2017.

\bibitem{tolias2015particular}
G.~Tolias, R.~Sicre, and H.~J{\'e}gou, ``Particular object retrieval with
  integral max-pooling of cnn activations,'' \emph{arXiv preprint
  arXiv:1511.05879}, 2015.

\bibitem{ISO2012_MPEGCDVS39219}
Y.~Lou, F.~Gao, Y.~Bai, J.~Lin, S.~Wang, J.~Chen, C.~Gan, V.~Chandrasekhar,
  L.~Duan, T.~Huang, and A.~Kot, ``{Improved retrieval and matching with CNN
  feature for CDVA},'' \emph{ISO/IEC JTC1/SC29/WG11/M39219}, Oct 2016.

\end{thebibliography}
% (used to reserve space for the reference number labels box

\end{document}